\documentclass[preprint, twocolumn]{elsarticle}
 \usepackage{setspace} 

\usepackage{hyperref}
\usepackage{cite}
\usepackage{amsmath,amssymb,amsfonts}
\usepackage{algorithmic}
\usepackage{graphicx}
\usepackage{textcomp}
\usepackage{arydshln}
\usepackage{multicol,lipsum}
\usepackage{subfig}  
\usepackage{multirow}
\usepackage{array}
\usepackage{soul}
\usepackage{lscape}
\usepackage{xcolor,colortbl}
\usepackage{booktabs}
\usepackage{soul}
\usepackage{longtable}
\usepackage{pdfpages}
\usepackage{float}
\usepackage{longtable}
\usepackage[skip=0pt]{caption}
\usepackage[ruled,vlined]{algorithm2e}

\journal{Nature Digital Medicine}

\newcolumntype{L}[1]{>{\raggedright\arraybackslash}p{#1}} 
\newcolumntype{C}[1]{>{\centering\arraybackslash}m{#1}} 
\newcolumntype{R}[1]{>{\raggedleft\arraybackslash}p{#1}} 

\newcommand{\specialcellC}[2][c]{%
\begin{tabular}[#1]{@{}c@{}}#2\end{tabular}}

\newcolumntype{P}[1]{>{\centering\arraybackslash}p{#1}}

\begin{document}

\onecolumn

\begin{frontmatter}
\title{The Medical Segmentation Decathlon}



\author[KCL]{Michela Antonelli\fnref{sharedFirst}\corref{mycorrespondingauthor}}
\cortext[mycorrespondingauthor]{Corresponding author}
\fntext[sharedFirst]{These authors contributed equally to this work}
\ead{michela.antonelli@kcl.ac.uk}
\author[CAMI,HIP,UniHDInf]{Annika Reinke\fnref{shared}}

\author[o1,o19,o20]{Spyridon Bakas}
\author[o2]{Keyvan Farahani}
\author[o3]{Annette Kopp-Schneider}
\author[o4]{Bennett A. Landman}
\author[o6]{Geert Litjens}
\author[o7]{Bjoern Menze}
\author[o8]{Olaf Ronneberger}
\author[o11]{Ronald M. Summers}
\author[o6]{Bram van Ginneken}


\author[o1]{Michel Bilello}
\author[o13]{Patrick Bilic} 
\author[o13]{Patrick F. Christ} 
\author[o14]{Richard K. G. Do} 
\author[o14]{Marc J. Gollub}
\author[o15]{Stephan H. Heckers} 
\author[o6]{Henkjan Huisman}
\author[o10]{William R. Jarnagin}
\author[o15]{Maureen K. McHugo} 
\author[o16]{Sandy Napel} 
\author[o14]{Jennifer S. Golia Pernicka}
\author[KCL]{Kawal Rhode}
\author[KCL]{Catalina Tobon-Gomez}
\author[o17]{Eugene Vorontsov} 

\author[o6]{Henkjan Huisman}
\author[o6]{James A. Meakin}
\author[KCL]{Sebastien Ourselin}
\author[o3]{Manuel Wiesenfarth}

\author[p5]{Pablo Arbel\'aez}
\author[p1]{Byeonguk Bae}
\author[p26]{Sihong	Chen}
\author[p5]{Laura Daza}
\author[p6]{Jianjiang Feng}
\author[p10]{Baochun He}
\author[p8]{Fabian Isensee}
\author[p9]{Yuanfeng Ji}
\author[p10]{Fucang	Jia}
\author[p11]{Namkug	Kim}
\author[p12]{Ildoo Kim}
\author[p13,p18]{Dorit Merhof}
\author[p14,p29]{Akshay Pai}
\author[p15]{Beomhee Park}
\author[p14]{Mathias Perslev}
\author[p17]{Ramin Rezaiifar}
\author[p13]{Oliver	Rippel}
\author[p19]{Ignacio Sarasua}
\author[p27]{Wei Shen}
\author[p1]{Jaemin Son}
\author[p19]{Christian Wachinger}
\author[p9]{Liansheng Wang}
\author[p28]{Yan Wang}
\author[p23]{Yingda	Xia}
\author[p24]{Daguang Xu}
\author[p6]{Zhanwei Xu}
\author[p26]{Yefeng	Zheng}

\author[o9]{Amber L. Simpson}
\author[CAMI,UniHDInf,UniHDMed,HIP]{Lena Maier-Hein\fnref{sharedLast}}
\fntext[sharedLast]{These authors contributed equally to this work}
\author[KCL]{M. Jorge Cardoso\fnref{sharedLast}}

\address[KCL]{School of Biomedical Engineering \& Imaging Sciences, King’s College London, London, UK}
\address[CAMI]{Div. Computer Assisted Medical Interventions, German Cancer Research Center (DKFZ), Heidelberg, Germany}
\address[HIP]{HIP Helmholtz Imaging Platform, German Cancer Research Center (DKFZ), Heidelberg, Germany}
\address[UniHDInf]{Faculty of Mathematics and Computer Science, University of Heidelberg, Heidelberg, Germany}
\address[o1]{Center for Biomedical Image Computing and Analytics (CBICA), University of Pennsylvania, Philadelphia, PA, USA}
\address[o2]{Center for Biomedical Informatics and Information Technology, National Cancer Institute (NIH), Bethesda, MD, USA}
\address[o3]{Div. Biostatistics, German Cancer Research Center (DKFZ), Heidelberg, Germany}
\address[o4]{Electrical Engineering and Computer Science, Vanderbilt University, Nashville TN, USA}

\address[o6]{Radboud University Medical Center, Radboud Institute for Health Sciences, Nijmegen, The Netherlands}
\address[o7]{Quantitative Biomedicine, University of Zurich, Zurich, Switzerland}
\address[o8]{DeepMind, London, UK}
\address[o9]{School of Computing/Department of Biomedical and Molecular Sciences, Queen’s University, Kingston, ON, Canada}
\address[o10]{Department of Surgery, Memorial Sloan Kettering Cancer Center, New York, NY, USA}
\address[o11]{Imaging Biomarkers and Computer-Aided Diagnosis Laboratory, Department of Radiology and Imaging Sciences, National Institutes of Health Clinical Center (NIH), Bethesda, MD, USA}

\address[o12]{Diagnostic Image Analysis Group, Radboud University Medical Center, Nijmegen, The Netherlands}
\address[o13]{Department of Informatics, Technische Universität München}
\address[o14]{Department of Radiology, Memorial Sloan Kettering Cancer Center, New York, NY, USA}
\address[o15]{Department of Psychiatry \& Behavioral Sciences, Vanderbilt University Medical Center, Nashville, TN, USA}
\address[o16]{Department of Radiology, Stanford University, Stanford, CA, USA}
\address[o17]{Department of Computer Science and Software Engineering, \'{E}cole Polytechnique de Montr\'{e}al}
\address[o19]{Department of Radiology, Perelman School of Medicine, University of Pennsylvania, Philadelphia, PA, USA}
\address[o20]{Department of Pathology and Laboratory Medicine, Perelman School of Medicine, University of Pennsylvania, Philadelphia, PA, USA}

\address[UniHDMed]{Medical Faculty, University of Heidelberg, Heidelberg, Germany}

\address[p1]{VUNO Inc., Seoul, Korea}
\address[p5]{Universidad de los Andes, Colombia}
\address[p6]{Department of Automation, Tsinghua University, Beijing, China}
\address[p8]{HIP Applied Computer Vision Lab, Division of Medical Image Computing, German Cancer Research Center (DKFZ), Heidelberg, Germany}
\address[p9]{Department of Computer Science, Xiamen University, Xiamen 361005, China}
\address[p10]{Shenzhen Institute of Advanced Technology, Chinese Academy of Sciences, Shenzhen, China}
\address[p11]{Affiliation of Participating teams}
\address[p12]{Kakao Brain, Republic of Korea}
\address[p13]{Institute of Imaging \& Computer Vision, RWTH Aachen University, Aachen, Germany}
\address[p29]{Cerebriu A/S, Copenhagen, Denmark}
\address[p14]{Department of Computer Science, University of Copenhagen,  Copenhagen, Denmark}
\address[p17]{MaaDoTaa.com, San Diego, California, USA}
\address[p18]{Fraunhofer Institute for Digital Medicine MEVIS, Bremen, Germany}
\address[p19]{Lab for Artificial Intelligence in Medical Imaging (AI-Med), Department of Child and Adolescent Psychiatry, University Hospital, LMU München, Germany}
\address[p23]{Johns Hopkins University, Baltimore, United States}
\address[p24]{NVIDIA, Santa Clara, CA, USA}
\address[p26]{ Tencent Jarvis Lab, Shenzhen, China}
\address[p27]{MoE Key Lab of Artificial Intelligence, AI Institute, Shanghai Jiao Tong University, Shanghai, China}
\address[p28]{Shanghai Key Laboratory of Multidimensional Information Processing, East China Normal University, Shanghai, China}

\begin{abstract}
International challenges have become the \textit{de facto} standard for comparative assessment of image analysis algorithms given a specific task. Segmentation is so far the most widely investigated medical image processing task, but the various segmentation challenges have typically been organized in isolation, such that algorithm development was driven by the need to tackle a single specific clinical problem. We hypothesized that a method capable of performing well on multiple tasks will generalize well to a previously unseen task and potentially outperform a custom-designed solution. To investigate the hypothesis, we organized the Medical Segmentation Decathlon (MSD)---a biomedical image analysis challenge, in which algorithms compete in a multitude of both tasks and modalities. The underlying data set was designed to explore the axis of difficulties typically encountered when dealing with medical images, such as small data sets, unbalanced labels, multi-site data and small objects. The MSD challenge confirmed that algorithms with a consistent good performance on a set of tasks preserved their good average performance on a different set of previously unseen tasks. Moreover, by monitoring the MSD winner for two years, we found that this algorithm continued generalizing well to a wide range of other clinical problems, further confirming our hypothesis. Three main conclusions can be drawn from this study: (1) state-of-the-art image segmentation algorithms are mature, accurate, and generalize well when retrained on unseen tasks; (2) consistent algorithmic performance across multiple tasks is a strong surrogate of algorithmic generalizability; (3) the training of accurate AI segmentation models is now commoditized to non AI experts.
\end{abstract}

\begin{keyword}
Image segmentation; Medical image; Deep learning; Grand challenge
\end{keyword}

\end{frontmatter}

\section{Introduction}
Machine learning is beginning to revolutionize many fields of medicine, with success stories ranging from the accurate diagnosis and staging of diseases~\citep{litjens2016deep}, to the early prediction of adverse events~\citep{poplin2018prediction} and the automatic discovery of antibiotics~\citep{stokes2020deep}. In this context, a large amount of literature has been dedicated to the automatic analysis of medical images~\citep{ayache201620th}.
Semantic segmentation refers to the process of transforming raw medical images into clinically relevant, spatially structured information, such as outlining tumor boundaries, and is an essential prerequisite for a number of clinical applications, such as radiotherapy planning~\citep{liang2019deep} and treatment response monitoring~\citep{assefa2010robust}. It is so far the most widely investigated medical image processing task, with about 70\% of all biomedical image analysis challenges dedicated to it~\citep{MaierHein2018}. With thousands of algorithms published in the field of biomedical image segmentation per year~\citep{isensee2021nnu}, however, it has become challenging to decide on a baseline architecture as starting point when designing an algorithm for a new given clinical problem. 

\begin{figure}
\centering
\includegraphics[width=1.2\textwidth]{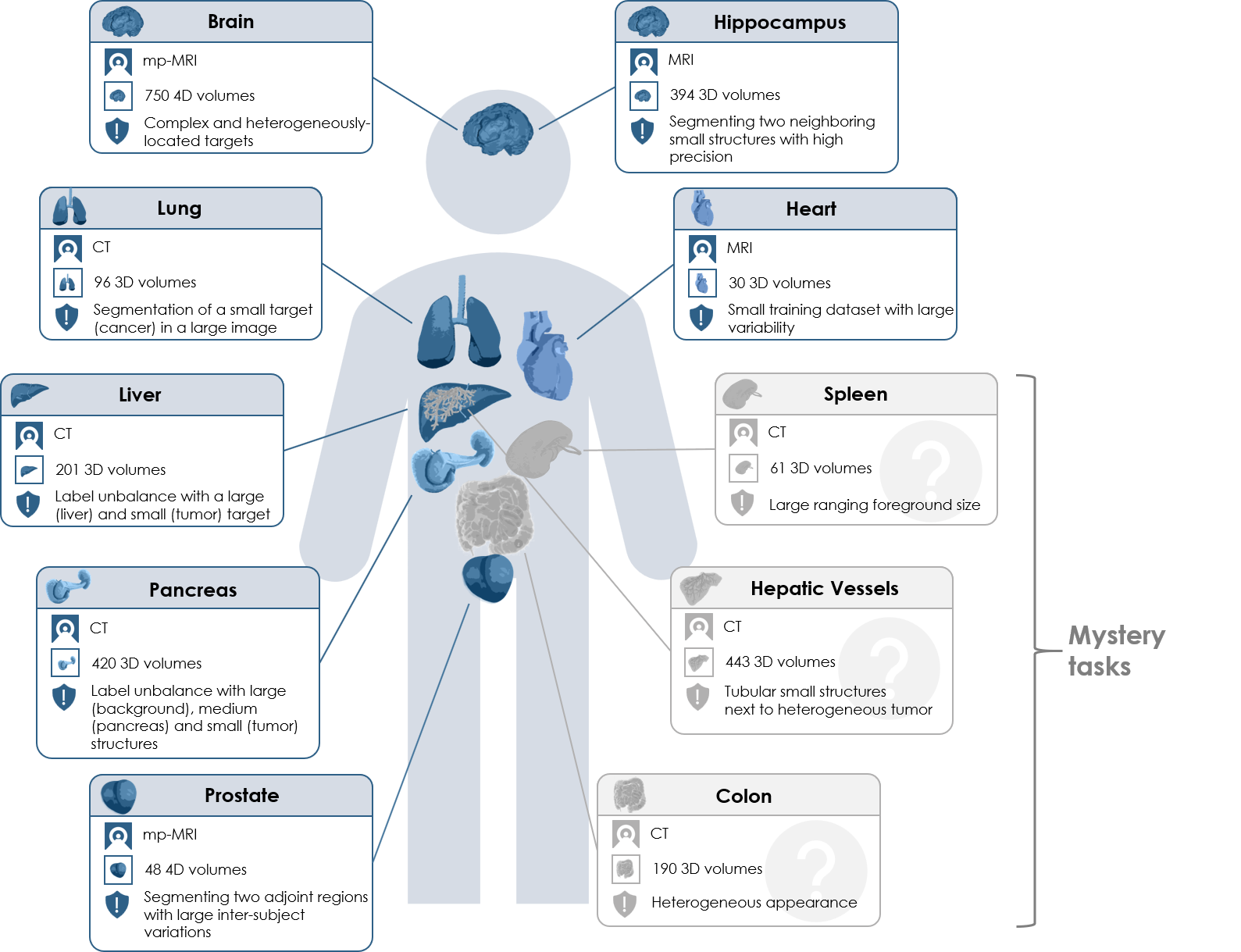}
\caption{Overview of the ten different tasks of the Medical Segmentation Decathlon (MSD). The challenge comprised different target regions, modalities and challenging characteristics and was separated into seven known tasks (blue; the development phase) and three mystery tasks (gray; the mystery phase). Used abbreviations: MRI —- magnetic resonance imaging, mp-MRI —- multiparametric-magnetic resonance imaging, CT —- computed tomography.}
\label{fig:overview}
\end{figure}

International challenges have become the de-facto standard for comparative assessment of image analysis algorithms given a specific task~\citep{MaierHein2018}. Yet, a deep learning architecture well-suitable for a certain clinical problem (e.g., segmentation of brain tumors) may not necessarily generalize well to different, unseen tasks (e.g., vessel segmentation in the liver). Such a \textit{"generalizable learner"}, which in this setting would represent a fully automated method that can learn any segmentation task given some training data and without the need for human intervention, would provide the missing technical scalability to allow many new applications in computer-aided diagnosis, biomarker extraction, surgical intervention planning, disease prognosis, etc. To address this gap in the literature, we proposed the concept of the \textit{Medical Segmentation Decathlon (MSD)}, an international challenge dedicated to identifying a general-purpose algorithm for medical image segmentation. The competition comprised ten different data sets with various challenging characteristics, as shown in Fig.~\ref{fig:overview}. The participants were allowed to submit only one solution, able to solve all problems without changing the architecture or hyperparameters.

The contribution of this paper is three-fold: (1) We are the first to organize a biomedical image analysis challenge in which algorithms compete in a multitude of both tasks and modalities. More specifically, the underlying data set has been designed to feature some of the representative difficulties typically encountered when dealing with medical images, such as small data sets, unbalanced labels, multi-site data and small objects. (2) Based on the MSD, we released the first open framework for benchmarking medical segmentation algorithms with a specific focus on generalizability. (3) By monitoring the winning algorithm, we show that generalization across various clinical applications is possible with one single framework.

In the following, we will first describe the challenge design, including the organization, mission, data sets and assessment method, in Section~\ref{sec:methods}, followed by the presentation of the results in Section~\ref{sec:results}, in which we present the submitted methods and rankings as well as the results for the live challenge. We conclude with a discussion in Section~\ref{sec:discussion}.

\section{Methods}
\label{sec:methods}
This section is organized according to the EQUATOR\footnote{https://www.equator-network.org} guideline BIAS (Bio\-medical Image Analysis ChallengeS) \citep{maier2020bias}, a recently published guideline specifically designed for the reporting of biomedical image analysis challenges. It comprises information on challenge organization and mission, as well as the data sets and assessment methods used to evaluate the submitted results. 

\subsection{Challenge organization}
The Decathlon challenge was organized at the International Conference on Medical Image Computing and Computer Assisted Intervention (MICCAI) 2018, held in Granada, Spain. After the main challenge event at MICCAI, a live challenge was opened for submissions which is still open and regularly receives new submissions (more than 500 as of May 15th 2021).

The MSD challenge aimed to test the ability of machine learning algorithms to accurately segment a large collection of prescribed regions of interest, as defined by ten different data sets, each corresponding to a different anatomical structure (see Fig.~\ref{fig:overview}) and to at least one medical imaging \textit{task}~\citep{2018Lena}. 
The challenge itself consisted of two phases:

In the first phase, named the \textit{development phase}, the training cases (comprising images and labels) for seven data sets were released, namely for brain, liver, heart, hippocampus, prostate, lung, and pancreas. Participants were expected to download the data, develop a general purpose learning algorithm, train the algorithm on each task's training data independently and without human interaction (no task-specific manual parameter settings), run the learned model on each task's test data, and submit the segmentation results. Each team was only allowed to make one submission per day to avoid model overfit, and the results were presented in form of a live leaderboard on the challenge website,\footnote{http://medicaldecathlon.com/} visible to the public. Note that participants were only able to see the average performance obtained by their algorithm on the test data of the seven development tasks. 

The purpose of the second phase of the challenge, named the \textit{mystery phase}, was to investigate whether algorithms were able to generalize to unseen segmentation tasks. Teams that submitted to the first phase and completed all necessary steps were invited to download three more data sets (images and labels), i.e., hepatic vessels, colon, and spleen. They were allowed to train their previously developed algorithm on the new data, without any modifications to the method itself. Segmentation results of the mystery phase could only be submitted once. A detailed description of the challenge organization is summarized in ~\ref{app:organization}, following the form introduced in \citep{maier2020bias}. 

\subsection{The Decathlon mission}
\label{subsec:missionchallenge}
Medical image segmentation, i.e., the act of labeling or contouring structures of interest in medical imaging data, is a task of crucial importance, both clinically and scientifically, as it allows the quantitative characterization of regions of interest. When performed by human raters, image segmentation is very time-consuming, thus limiting its clinical usage. Algorithms can be used to automate this segmentation process, but, classically, a different algorithm had to be developed for each segmentation task. The goal of the MSD challenge was finding a single algorithm, or learning system, that would be able to generalize and work accurately across multiple different medical segmentation tasks, without the need for any human interaction.

The tasks of the Decathlon challenge were chosen as a representative sample of real-world applications, so as to test for algorithmic generalizability to these. Different axes of complexity were explicitly explored: the type and number of input modalities, the number of regions of interest, their shape and size, and the complexity of the surrounding tissue environment (see Fig.~\ref{fig:overview}). Detailed information of each data set is provided in Section~\ref{subsec:data sets} and Table~\ref{tabData}.

\subsection{Challenge data sets}
\label{subsec:data sets}

The Decathlon challenge made ten data sets available online \citep{AmberPaper}, where each data set had between one and three region-of-interest (ROI) targets (17 targets in total). Importantly, all data sets have been released with a permissive copyright-license (CC-BY-SA 4.0), thus allowing for data sharing, redistribution, and commercial usage, and subsequently promoting the data set as a standard test-bed for all users. 
The images (2,633 in total) were acquired across multiple institutions, anatomies and modalities during real-world clinical applications. All images were de-identified and reformatted to the Neuroimaging Informatics Technology Initiative (NIfTI) format \url{https://nifti.nimh.nih.gov}. All images were transposed (without resampling) to the most approximate right-anterior-superior coordinate frame, ensuring the data matrix $x$-$y$-$z$ direction was consistent. Lastly, non-quantitative modalities (e.g., MRI) were robust min-max scaled to the same range. 
For each segmentation task, a pixel-level label annotation was provided depending on the definition of each specific task. For 8 out of 10 data sets, two thirds of the data were released as training sets (images and labels) and one third as test set (images without labels). As the remaining two tasks (brain tumor and liver) consisted of data from two well-known challenges, the original training/test split was preserved.  
 
Table \ref{tabData} presents a summary of the ten data sets, including the modality, image series, ROI targets and data set size. A brief description of each data set is provided below.

 
\begin{itemize}
\item \textbf{Development Phase (1st)} contained seven data sets with thirteen target regions-of-interest in total:

\begin{itemize}
\item \textbf{Brain}: The data set consists of 750 multiparametric magnetic resonance images (mp-MRI) from patients diagnosed with either glioblastoma or lower-grade glioma. The sequences used were native T1-weighted (T1), post-Gadolinium (Gd) contrast T1-weighted (T1-Gd), native T2-weighted (T2), and T2 Fluid-Attenuated Inversion Recovery (FLAIR). The corresponding target ROIs were the three tumor sub-regions, namely edema, enhancing, and non-enhancing tumor. This data set was selected due to the challenge of locating these complex and heterogeneously-located targets.
The data was acquired from 19 different institutions and contained a subset of the data used in the 2016 and 2017 Brain Tumor Segmentation (BraTS) challenges \citep{brats, bakas2017advancing, bakas2018identifying}.
\item \textbf{Heart}: The data set consists of 30 mono-modal MRI scans of the entire heart acquired during a single cardiac phase (free breathing with respiratory and electrocardiogram (ECG) gating). The corresponding target ROI was the left atrium.
This data set was selected due to the combination of a small training data set with large anatomical variability. 
The data was acquired as part of the 2013 Left Atrial Segmentation Challenge (LASC) \citep{heart}.
\item \textbf{Hippocampus}: The data set consists of 195 MRI images acquired from 90 healthy adults and 105 adults with a non-affective psychotic disorder. T1-weighted MPRAGE was used as the imaging sequence. The corresponding target ROIs were the anterior and posterior of the hippocampus, defined as the hippocampus proper and parts of the subiculum. This data set was selected due to the precision needed to segment such a small object in the presence of a complex surrounding environment. 
The data was acquired at the Vanderbilt University Medical Center, Nashville, US. 
\item \textbf{Liver}: The data set consists of 201 contrast-enhanced CT images from patients with primary cancers and metastatic liver disease, as a consequence of colorectal, breast, and lung primary cancers. The corresponding target ROIs were the segmentation of the liver and tumors inside the liver. This data set was selected due to the challenging nature of having significant label unbalance between large (liver) and small (tumor) target region of interests (ROIs).
The data was acquired in the IRCAD Hôpitaux Universitaires, Strasbourg, France and contained a subset of patients from the 2017 Liver Tumor Segmentation (LiTS) challenge \citep{liver}.
\item \textbf{Lung}: The data set consists of preoperative thin-section CT scans from 96 patients with non-small cell lung cancer. The corresponding target ROI was the tumors within the lung. This data set was selected due to the challenge of segmenting small regions (tumor) in an image with a large field-of-view. 
Data was acquired via the Cancer Imaging.Archive\footnote{https://www.cancerimagingarchive.net/}
\item \textbf{Prostate}: The data set consists of 48 prostate multiparametric MRI (mpMRI) studies comprising T2-weighted, Diffusion-weighted and T1-weighted contrast enhanced series. A subset of two series, transverse T2-weighted  and the apparent diffusion coefficient (ADC) was selected. The corresponding target ROIs were the prostate peripheral zone (PZ) and the transition zone (TZ). This data set was selected due to the challenge of segmenting two adjoined regions with very large inter-subject variability. 
The data was acquired at Radboud University Medical Center, Nijmegen Medical Centre, Nijmegen, The Netherlands. 
\item \textbf{Pancreas}: The data set consists of 421 portal-venous phase CT scans of patients undergoing resection of pancreatic masses. The corresponding target ROIs were the pancreatic parenchyma and pancreatic mass (cyst or tumor). This data set was selected due to label unbalance between large (background), medium (pancreas) and small (tumor) structures.
The data was acquired in the Memorial Sloan Kettering Cancer Center, New York, US.
\end{itemize}

\item \textbf{Mystery Phase (2nd)} contained three (hidden) data sets with four target regions-of-interest in total:
\begin{itemize}
\item \textbf{Colon}: The data set consists of 190 portal venous phase CT scans of patients undergoing resection of primary colon cancer. The corresponding target ROI was colon cancer primaries. This data set was selected due to the challenge of the heterogeneous appearance, and the annotation difficulties. 
The data was acquired in the Memorial Sloan Kettering Cancer Center, New York, US.
\item \textbf{Hepatic Vessels}: The data set consists of 443 portal venous phase CT scans obtained from patients with a variety of primary and metastatic liver tumors. The corresponding target ROIs were the vessels and tumors within the liver. This data set was selected due to the tubular and connected nature of hepatic vessels neighboring heterogeneous tumors. The data was acquired in the Memorial Sloan Kettering Cancer Center, New York, US.
\item \textbf{Spleen}: The data set consists of 61 portal venous phase CT scans from patients undergoing chemotherapy treatment for liver metastases. The corresponding target ROI was the spleen. This data set was selected due to the large variations in the field-of-view. The data was acquired in the Memorial Sloan Kettering Cancer Center, New York, US.
\end{itemize}
\end{itemize}

\begin{table}[H]
\caption{Summary of the ten data sets of the Medical Segmentation Decathlon. Used abbreviations: mp-MRI---multiparametric-magnetic resonance imaging, FLAIR---fluid-attenuated inversion recovery, T1w---T1 weighted image, T1 \textbackslash w Gd---post-Gadolinium (Gd) contrast T1-weighted image, T2w---T2 weighted image, CT---computed tomography, PZ---peripheral zone, TZ---transition zone.}
\label{tabData}
\centering
\scalebox{0.9}{
\begin{tabular}{p{0.3cm} C{1.9cm} C{1.2cm} C{2.5cm} C{4.1cm} C{3.4cm} }
\toprule
\textbf{\scriptsize{Phase}}&\textbf{\scriptsize{Task}}& \textbf{\scriptsize{Modality}} &\textbf{{\specialcellC{\scriptsize{Protocol}}}}& \textbf{\scriptsize{Target}}& \textbf{\specialcellC{\scriptsize{\# Cases (Train/Test)}}} \\ \midrule
\parbox[t]{1mm}{\multirow{9}{*}{\rotatebox[origin=c]{90}{\textbf{\tiny{Development phase}}}}} 
&\scriptsize{Brain} & \scriptsize{mp-MRI} & \scriptsize{FLAIR, T1w, T1 \textbackslash w Gd, T2w} & \scriptsize{Edema, enhancing and non-enhancing tumor} & \scriptsize{750 4D volumes (484/266)} \\

&\scriptsize{Heart} & \scriptsize{MRI} & \scriptsize{---} & \scriptsize{Left atrium} & \scriptsize{30 3D volumes (20/10)}  \\

&\scriptsize{Hippocampus} & \scriptsize{MRI} & \scriptsize{T1w} & \scriptsize{Anterior and posterior of hippocampus} & \scriptsize{394 3D volumes (263/131)}  \\

&\scriptsize{Liver} & \scriptsize{CT} & \scriptsize{Portal venous phase}& \scriptsize{Liver and liver tumor}  & \scriptsize{210 3D volumes (131/70)}  \\

&\scriptsize{Lung} & \scriptsize{CT} & \scriptsize{---} & \scriptsize{Lung and lung cancer} & \scriptsize{96 3D volumes (64/32)}  \\

&\scriptsize{Pancreas} & \scriptsize{CT} & \scriptsize{Portal venous phase} & \scriptsize{Pancreas and pancreatic tumor mass} & \scriptsize{420 3D volumes (282/139)} \\

&\scriptsize{Prostate} & \scriptsize{mp-MRI} & \scriptsize{T2, ADC} & \scriptsize{Prostate PZ and TZ} & \scriptsize{48 4D volumes (32/16)} \\
\midrule
\parbox[t]{1mm}{\multirow{5}{*}{\rotatebox[origin=c]{90}{\textbf{\tiny{Mystery phase}}}}}&&&&&\\
& \scriptsize{Colon} & \scriptsize{CT} & \scriptsize{Portal venous phase} & \scriptsize{Colon cancer primaries} & \scriptsize{190 3D volumes (126/64)} \\

&\scriptsize{Hepatic Vessels} & \scriptsize{CT} & \scriptsize{Portal venous phase} & \scriptsize{Hepatic vessels and hepatic tumor} & \scriptsize{443 3D volumes (303/140)}  \\

&\scriptsize{Spleen} & \scriptsize{CT} & \scriptsize{Portal venous phase} & \scriptsize{spleen} & \scriptsize{61 3D volumes (41/20)}  \\
\bottomrule

\end{tabular}}
\end{table}

\subsection{Assessment method}
\label{subsec:assessmentmethods}

\subsubsection{Assessment of competing teams}
\label{sec:ranking}
Two widely known semantic segmentation metrics were used to evaluate the submitted approaches, namely the \textit{Dice Similarity Coefficient (DSC)} \citep{dice1945measures} and the \textit{Normalised Surface Distance (NSD)} \citep{nikolov2018deep}, both computed on 3D volumes. The implementation of both metrics can be downloaded in the form of a Jupyter notebook from the challenge website,\footnote{http://www.medicaldecathlon.com section \textit{Assessment Criteria}}. The metrics \(DSC and NSD\) were chosen due to their popularity, rank stability \citep{2018Lena}, and smooth, well-understood and well-defined behavior when ROIs do not overlap. Having simple and rank-stable metrics also allows the statistical comparison between methods. It is important to note that the proposed metrics are not task-specific nor task-optimal, and thus, they do not fulfill the necessary criteria for clinical algorithmic validation of each task, as discussed in Section~\ref{subsec:dis-assessment}. 

A so-called significance score was determined for each algorithm $a$, separately for each task/target ROI $c_i$ and metric $m_j \in \{DSC, NSD\}$ and referred to as $s_{i,j}(a)$. Similarly to what was used to infer the ranking across the different BRATS tasks \citep{Bakas_et_al}, the significance score was computed according to the following four-step process:
\begin{enumerate}
\item \textit{Performance assessment per case}: Determine performance $m_j(a_l,t_{ik})$ of all algorithms $a_l$, with $l=\{1, \ldots, N_A\}$, for all test cases $t_{ik}$, with $k=\{1, \ldots, N_i\}$, where $N_A$ is the number of competing algorithms and $N_i$ is the number of test cases in competition $c_i$. Set $m_j(a_l,t_{ik})$ to 0 if its value is undefined.
\item \textit{Statistical tests}: Perform a Wilcoxon signed-rank pairwise statistical test between algorithms $(a_l, a_{l'})$, with values $m_j(a_l, t_{ik}) - m_j(a_{l'}, t_{ik})$, $\forall k=\{1,...,N_i\}$.
\item \textit{Significance scoring}: $s_{i,j}(a_l)$ then equals the number of algorithms performing significantly worse than $a_l$, according to the statistical test (per comparison $\alpha=0.05$, not adjusted for multiplicity).
\item \textit{Significance ranking}: The ranking is computed from the scores $s_{i,j}(a_l)$, with the highest score (rank 1) corresponding to the best algorithm. Note that shared scores/ranks are possible. If a task has multiple target ROI, the ranking scheme is applied to each ROI separately, and the final ranking per task is computed as the mean significance rank.
\end{enumerate}
The final score for each algorithm over all tasks of the development phase (the seven development tasks) and over all tasks of the mystery phase (the three mystery tasks) was computed as the average of the respective task's significance ranks. The full validation algorithm was defined and released prior to the start of the challenge, and available on the decathlon website.\footnote{http://medicaldecathlon.com/files/MSD-Ranking-scheme.pdf}

To investigate ranking uncertainty and stability, bootstrapping methods were applied with 1,000 bootstrap samples as described in \citep{2018Lena}. The statistical analysis was performed using the open-source R toolkit \textit{challengeR}\footnote{https://phabricator.mitk.org/source/challenger/}, version 1.0.1 \citep{wiesenfarth2021methods}, for analyzing and visualizing challenge results . The original rankings computed for the development and mystery phases were compared to the ranking lists based on the individual bootstrap samples. The correlation of pairwise rankings was determined via Kendall's $\tau$ \citep{kendall1938new}, which provides values between $-1$ (for reverse ranking order) and $1$ (for identical ranking order). 

\subsubsection{Monitoring of the challenge winner and algorithmic progress} 
To investigate our hypothesis that a method capable of performing well on multiple tasks will generalize its performance to an unseen task, and potentially even outperform a custom-designed task-specific solution, we monitored the winner of the challenge for a period of two years. Specifically, we reviewed the rank analysis and leaderboards presented in the corresponding article \citep{isensee2021nnu}, as well as the leaderboard of challenges from the grand-challenge.org website organized in 2020. We also reviewed further articles mentioning the new state-of-the-art method nnU-Net \citep{ma2021cutting}.
Finally, as the MSD challenge submission was reopened after the challenge event (denoted the "MSD Live Challenge"), we monitored submissions for new algorithmic approaches which achieve state-of-the-art performance, in order to probe new areas of scientific interest and development. 

\section{Results}
\label{sec:results}
\subsection{Challenge submissions}
\begin{figure}
\centering
\includegraphics[width=0.49\textwidth] {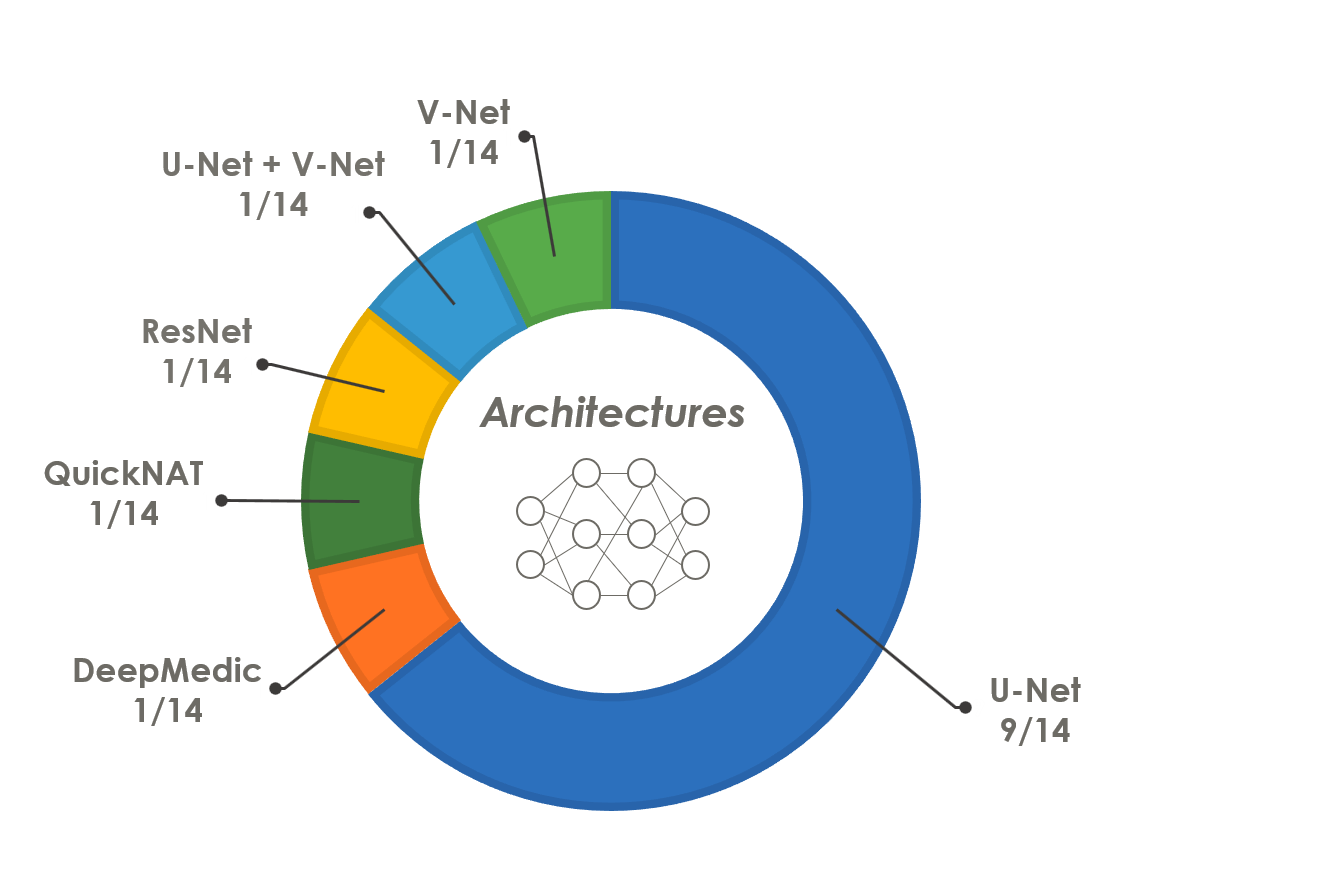}
\includegraphics[width=0.49\textwidth] {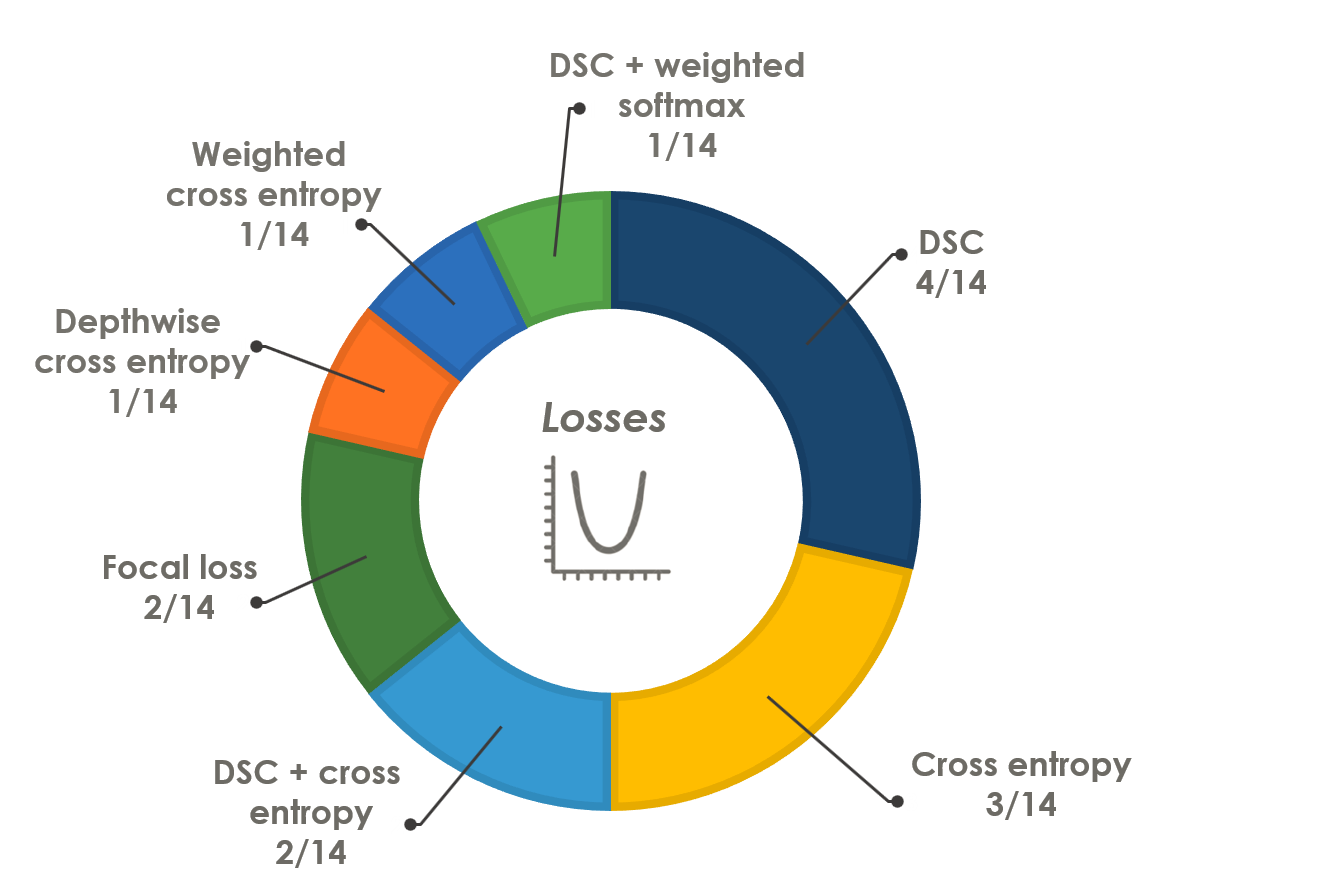}
\caption{Base network architectures (left) and loss functions (right) used by the participants of the 2018 Decathlon challenge who provided full algorithmic information.}
\label{statistics}
\end{figure}

In total, 180 teams registered for the challenge, from which 31 submitted fully-valid and complete results for the development phase. From these, a subset of 19 teams submitted final and valid results for the mystery phase. 
Among the methods that fulfilled all the criteria to move to the mystery phase, all methods were based on convolutional neural networks (CNNs), with the U-Net \citep{ronneberger2015u} being the most frequently used base architecture---employed by more than half of the teams (64\%). The most commonly used loss function was the DSC loss (29\%), followed by the cross entropy loss (21\%). Fig.~\ref{statistics} provides a complete list of both network architectures and loss functions used in the challenge.  61\% of the teams used the adaptive moment estimation (Adam) optimizer \citep{kingma2014adam}, while the stochastic gradient descent (SGD) \citep{zhang2004solving} was used by 33\% of the teams. 

\begin{landscape}
\begin{table}[htbp]
  \centering
  \tiny
   \caption{Details of the participating teams' methods.}
   \label{methodDes}

  \setlength\tabcolsep{3pt} 
    \begin{tabular}{m{1.2cm}>{\raggedright}m{3.2cm}>{\raggedright}m{2.4cm}>{\raggedright}m{1.9cm}>{\raggedright}m{2.8cm}>{\raggedright}m{3cm}>{\raggedright}m{1.8cm} m{2.6cm}
    }
 \toprule
    \textbf{Team} &\textbf{Base Architecture \&  Modifications} & \textbf{Augmentation Strategy}&\textbf{Loss Function} & \textbf{Optimizer Training} &\textbf{Pre-processing} &\textbf{Post-processing} & \textbf{Ensembling} \textbf{strategy} \\
     \midrule
     
    nnU-Net & U-Net---leaky ReLU, instance normalization, strided convolutions& Affine, non-linear, intensity, flipping along all axes, random crop & \specialcellC{DSC loss \\ cross entropy loss} & Adam (lr$=$3e-4, \\ weight decay$=$3e-5) \\ 1,000 epochs with early stopping & Intensity normalization, padding/cropping, rescaling interpolation & Segmentation region removal & Training set cross-validation model selection \\
     \midrule
     NVDLMED & ResNet, anisotropic 3D kernels& Affine,left-right flip, random crop & DSC loss  & SGD (lr$=7$e-3, \\ weight decay$=$3e-5, momentum$=0.9$, batch size$=8$) \\40,000 epochs & Intensity normalization, cropping, rescaling interpolation & Region removal & Ensemble three trained views \\
     \midrule
    K.A.V.athlon & U-Net, V-Net Squeeze-and-Excitation & Affine, noise, left-right flip, random crop, blur & DSC loss & Adam (lr$=$1e-4)  & Intensity normalization, rescaling interpolation & Segmentation region removal & ---\\
     \midrule
    LS Wang's Group & U-Net residual blocks nested with dilations & Affine, histogram, left-right flip, random crop & Modified focal loss  & Adam (lr=1e-4), 90,000 epochs  & Padding/cropping, rescaling interpolation &---& Averaging multiple runs\\
     \midrule
    MIMI & U-Net skip connections inside convolutional blocks, leaky ReLU, dropout added inside decoders & Affine, noise, left-right flip, random crop & DSC loss  & Adam (lr$=$0.5e-4)  & Intensity normalization, rescaling interpolation & Intensity-based region removal &Averaging multiple runs\\
     \midrule
    Cerebriu DIKU & U-Net, batch normalization layers & Non-linear, multi-planar sampling & Cross entropy loss & Adam (lr$=$5e-5, $\beta_1=$0.9, $\beta_2=0.999$, $\epsilon=$1e-8)  & Intensity normalization, multi-planar sampling  & -- & Averaging multiple runs\\
     \midrule
    Whale  & U-Net 3D/2D & Left-right flip, random crop & Cross entropy loss & SGD (lr$=$1e-3, momentum$=0.99$)  & Padding/cropping & Segmentation region removal and masking & Multiple-architecture non-weighted averaging \\
     \midrule
    Lupin  & U-Net & Affine, noise, random crop & Focal loss  & Adam (lr$=$1e-4)  & Intensity normalization & -- &---\\
     \midrule
    SIAT\_MIDS  & V-Net  batch normalization & Non-linear, histogram & DSC and weighted softmax loss  & SGD (lr$=$1e-4/0.5e-4)  & Intensity normalization, padding/cropping, rescaling interpolation & Segmentation region removal &---\\
     \midrule
  LfB  & U-Net residual connections, instance normalization & Affine, noise, non-linear, random crop & DSC loss  & Adam (lr$=$0.5e-4), 30,000 epochs  & Padding/cropping, rescaling interpolation, resampling to the median voxel-spacing &---&---\\
     \midrule
    VST  & U-Net 2.5D/3D & Affine & Cross entropy loss & Adam and SGD (lr$=0.1$, momentum$=0.9$)  & Intensity normalization, padding/cropping, rescaling interpolation & Segmentation region removal and intensity-based masking &Multiple-architecture weighted averaging \\
     \midrule
    AI-MED & QuickNAT CRF &---& DSC loss, cross entropy loss & SGD (lr$=0.01$, batch size $=4$, momentum$=0.9$), 20 epochs &  Padding/cropping & Conditional random field/graph-cuts refinement &---\\
     \midrule
    Lesswire1  & U-Net  & Test-time augmentation & Depth-wise cross-entropy loss & Adam (lr$=$1e-4)  & Intensity normalization &---&---\\
     \midrule
    BCV Uniandes  & DeepMedic &---&Cross entropy loss & Adam (lr$=$1e-4)  & Intensity normalization, rescaling interpolation & Region removal &---\\
     \bottomrule
    \end{tabular}%
\end{table}%
\end{landscape}

\subsection{Method description of top three algorithms}
\label{subsec:methoddescription}
In the following, the top three methods are briefly described. Table~\ref{methodDes} provides an overview over all methods that were submitted for the mystery phase. For more details, see \ref{app:methods}.

\subsubsection{nnU-Net}
The key idea of \textit{nnU-Net's }method was to use a fully automated dynamic adaptation of the segmentation pipeline, done independently for each task in the MSD, based on an analysis of the respective training data set. Image pre-processing, network topologies and post-processing were determined fully automatically and considered more important than the actual architecture ~\citep{isensee2021nnu}. 
\textit{nnU-Net} was based on the U-Net architecture \citep{ronneberger2015u} with the following modifications: the use of leaky ReLU, instance normalization and strided convolutions for downsampling~\citep{isensee2021nnu}. It further applied a combination of augmentation strategies, namely affine transformation, non-linear deformation, intensity transformation (similar to gamma correction), mirroring along all axes and random crop. The sum of the DSC and cross entropy loss was used, while utilizing the Adam optimizer. 
The method applied a purposely defined ensembling strategy in which four different architectures were used. The selection of the task-specific optimal combination was found automatically via cross-validation on the training set. 

\subsubsection{NVDLMED}
The key idea of \textit{NVDLMED's} method was to use a fully-supervised uncertainty-aware multi-view co-training strategy \citep{xia20203d}. They achieved robustness and generalization by initializing the model from 2D pre-trained models and using three views of networks to gain more 3D information through the multi-view co-training process. 
They further used a resampling strategy to cope with the differences among the ten tasks.
The \textit{NVDLMED} team utilized a 3D version of the ResNet \citep{he2016deep} with anisotropic 3D kernels~\citep{xia20203d}. The team further applied a combination of augmentation strategies, namely affine transformation, geometric left-right flip and random crop. The DSC loss and the SGD optimizer were employed. 
\textit{NVDLMED} ensembled three models, each trained on a different view (coronal, saggital and axial).

\subsubsection{K.A.V.athlon}
The key idea of \textit{K.A.V.athlon's} method was a generalization strategy in the spirit of AutoML \citep{he2021automl}. The process was designed to train and predict automatically using given image data and description without any parameter change or intervention by a human.
\textit{K.A.V.athlon's} method was based on a combination of the V-Net and U-Net architectures with the addition of a Squeeze-and-Excitation (SE) block and a residual block. The team further applied different types of augmentation, namely affine transformation, noise application, geometric left-right flip, random crop, and blurring. The DSC loss with a thresholded ReLU (threshold 0.5) and the Adam optimizer were employed. No ensembling strategy was used. 

\subsection{Individual performances and rankings}
\begin{figure}[H]
    \makebox[\linewidth]{
        \includegraphics[width=1\linewidth]{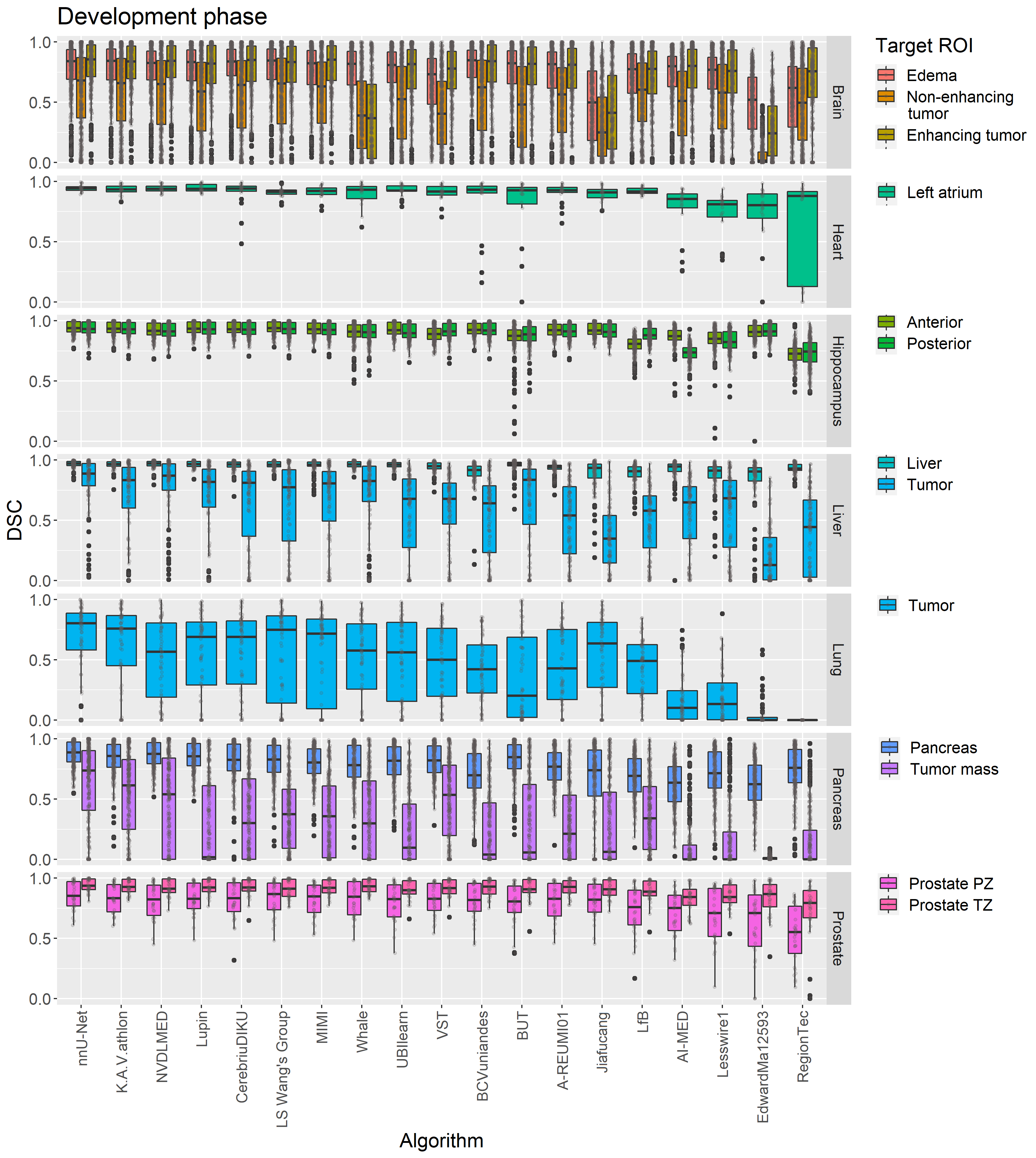}
    }
    \caption{Dot- and box-plots of the DSC values of all participating algorithms for the seven tasks of the development phase, color-coded by the target regions. box-plots represent descriptive statistics over all test cases. The median value is shown by the black horizontal line within the box, the first and third quartiles as the lower and upper border of the box, respectively, and the 1.5 interquartile range by the vertical black lines. Outliers are shown as black circles. The raw DSC values are provided as gray circles. Used abbreviations: PZ---peripheral zone, TZ---transition zone.}
    \label{fig:dscbox-plotsph1}
\end{figure}

\begin{figure}[H]
    \makebox[\linewidth]{
        \includegraphics[width=1\linewidth]{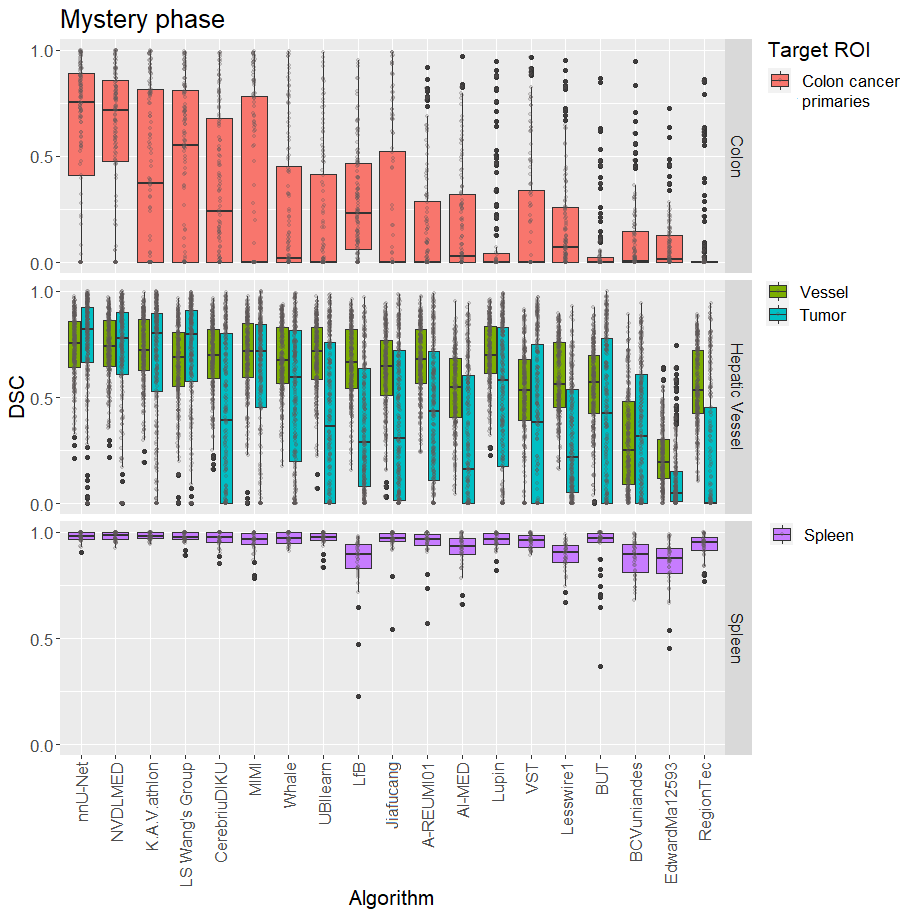}
    }
    \caption{Dot- and box-plots of the DSC values of all participating algorithms for the three tasks of the mystery phase, color-coded by the target regions. box-plots represent descriptive statistics over all test cases. The median value is shown by the black horizontal line within the box, the first and third quartiles as the lower and upper border of the box, respectively, and the 1.5 interquartile range by the vertical black lines. Outliers are shown as black circles. The raw DSC values are provided as gray circles.}
    \label{fig:dscbox-plotsph2}
\end{figure}

The DSC values for all participants for the development phase and the mystery phase are provided as box-plots in Figs.~\ref{fig:dscbox-plotsph1} and~\ref{fig:dscbox-plotsph2}, respectively. For tasks with multiple target ROIs (e.g., edema, non-enhancing tumor and enhancing tumor segmentation for the brain data set), the box-plots were color-coded according to the ROI. The distribution of the NSD metric values was comparable to the DSC values. 

It can be seen that the performance of the algorithms as well as their robustness depends crucially on the task and target ROI. The median of the mean DSC computed considering all test cases of a single task over all participants ranged from 0.16 (colon cancer segmentation (the mystery phase), cf. Table~\ref{tab:meanDSC-subtasks-colon-ph2}) to 0.94 (liver (the development phase), cf. Table~\ref{tab:meanDSC-subtasks-liver-ph1} and spleen segmentations (the mystery phase), cf. Table~\ref{tab:meanDSC-subtasks-spleen-ph2}). The full list of values are provided in~\ref{app:dsc-subtasks}.

\begin{figure}[H]
\vspace{-1.5cm}
     \centering
    \includegraphics[width=\textwidth]{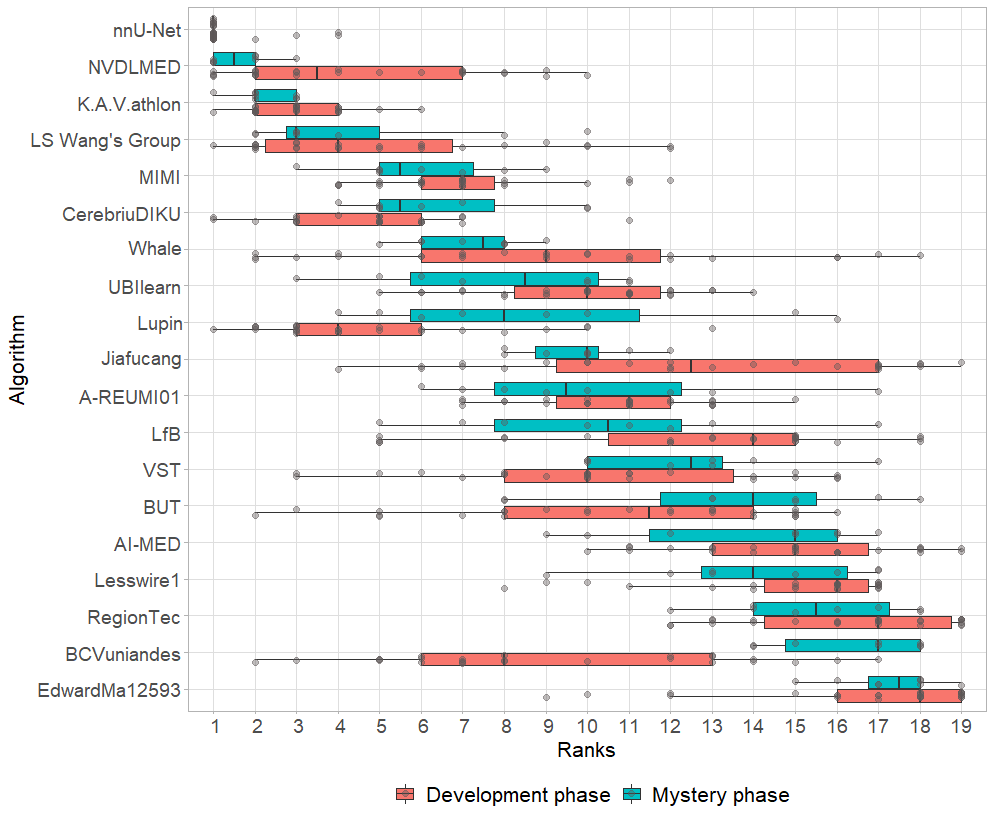}
     \caption{box-plots of ranks for all participating algorithms over all seven tasks and thirteen target regions of the development phase (red) and all three tasks and four target regions of the mystery phase (blue). The median value is shown by the black vertical line within the box, the first and third quartiles as the lower and upper border of the box, respectively, and the 1.5 interquartile range by the horizontal black lines. Individual ranks are shown as gray circles.}
     \label{fig:rankbox-plots}
\end{figure}

The rankings for the challenge are shown in Table~\ref{tab:ranking}. The winning method (\textit{nnU-Net}) was extremely robust with respect to the different tasks and target regions for both phases (cf. Figs.~\ref{fig:dscbox-plotsph1} and~\ref{fig:dscbox-plotsph2}). Ranks 2 and 3 switched places (\textit{K.A.V.athlon} and \textit{NVDLMED}) for both the development and mystery phase. Fig.~\ref{fig:rankbox-plots} further shows the ranks of all algorithms for all thirteen target regions of the development phase (red) and all four target regions of the mystery phase in form of a box-plot. Many teams show a large variation in their ranks across target ROIs. The lowest rank difference of three ranks was achieved for team \textit{nnU-Net} (minimum rank: 1, maximum rank: 4; the development phase) and the largest rank difference of sixteen ranks is obtained for team \textit{Whale} (minimum rank: 2, maximum rank: 18; the development phase). 

To investigate ranking robustness for different ranking methods, line plots \citep{wiesenfarth2021methods} are provided in \ref{app:dsc-subtasks} for all individual target regions. Furthermore, a comparison of the achieved ranks of algorithms for 1,000 bootstrapped samples is provided in the form of a stacked frequency plot \citep{wiesenfarth2021methods} in Fig.~\ref{fig:stackedplots}. For each participant, the frequency of the achieved ranks is provided for every task individually. It can be easily seen from both uncertainty analyses that team \textit{nnU-Net} implemented an extremely successful method that was at rank 1 for nearly every tasks and bootstrap sets.

The agreement of the original rankings computed for the development phase and the mystery phase and the ranking lists based on the individual bootstrap samples was determined via Kendall's $\tau$. The median (interquartile range (IQR)) Kendall's $\tau$ was $0.94$ ($0.91,0.95$) for the colon task, $0.99$ ($0.98,0.99$) for the hepatic vessel task and $0.92$ ($0.89,0.94$) for the spleen task. This shows that the rankings for the mystery phase were stable against small perturbations.
\newpage
\begin{table}[H]
\vspace{-1.5cm}
\caption{Rankings for the development phase and the mystery phase, median and interquartile range (IQR) of the Dice Similarity Coefficient (DSC) values of all team. The ranking was computed as described in Section~\ref{sec:ranking}.}
\label{tab:ranking}
\centering
\small
\begin{tabular}{c c c c | c c c c}
\toprule
\multicolumn{4}{c}{\textbf{The development phase}} & \multicolumn{4}{c}{\textbf{The mystery phase}}\\
\textbf{Rank} & \textbf{Team ID} & \textbf{Median DSC} & \textbf{IQR DSC} & \textbf{Rank} & \textbf{Team ID} & \textbf{Median DSC} & \textbf{IQR DSC}\\
\midrule
1 & \textbf{\textit{nnU-Net}}& 0.79 & (0.61,0.88) & 1 & \textbf{\textit{nnU-Net}}& 0.71 & (0.58,0.82)\\
2 & \textit{K.A.V.athlon}& 0.77 & (0.58,0.86) &  2& \textit{NVDLMED}& 0.69 & (0.54,0.79)\\
3 & \textit{NVDLMED} & 0.78 & (0.57,0.87) & 3& \textit{K.A.V.athlon}& 0.67 & (0.49,0.79)\\  
4 & \textit{Lupin} & 0.75 & (0.52,0.86) & 4 &\textit{LS Wang's Group} & 0.64 & (0.46,0.78)\\
5 & \textit{CerebriuDIKU} & 0.76 & (0.52,0.88) & 5& \textit{CerebriuDIKU}& 0.56 & (0.15,0.71)\\
6 & \textit{LS Wang's Group }& 0.75 & (0.51,0.88) & 6& \textit{MIMI}& 0.65 & (0.45,0.75)\\
7 & \textit{MIMI} & 0.73 & (0.51,0.86) &7 & \textit{Whale} & 0.55 & (0.20,0.68)\\
8 & \textit{Whale} & 0.65 & (0.28,0.83) & 8& \textit{UBIlearn}&0.55 & (0.05,0.69)\\
9 & \textit{UBIlearn}& 0.72 & (0.40,0.85)  & 9& \textit{LfB}& 0.49 & (0.16,0.63)\\
10 & \textit{VST}& 0.69 & (0.39,0.84) & 10& \textit{Jiafucang}& 0.48 & (0.04,0.67)\\
11 & \textit{BCVuniandes} & 0.70 & (0.42,0.86) & 11& \textit{A-REUMI01} &  0.51 & (0.14,0.65)\\
12.5 & \textit{BUT}& 0.72 & (0.39,0.84) & 12& \textit{AI-MED} & 0.33 & (0.01,0.52)\\
12.5 & \textit{A-REUMI01}& 0.70 & (0.42,0.85) & 13& \textit{Lupin}& 0.57 & (0.19,0.69) \\
14 & \textit{Jiafucang} & 0.49 & (0.11,0.81) & 14& \textit{VST}& 0.41 & (0.00,0.64)\\
15 & \textit{LfB} & 0.68 & (0.43,0.82) & 15& \textit{Lesswire1}& 0.40 & (0.08,0.52)\\
16 & \textit{AI-Med}& 0.63 & (0.29,0.79) & 16& \textit{BUT}& 0.38 & (0.01,0.60)\\
17 & \textit{Lesswire1}& 0.65 & (0.33,0.79) & 17& \textit{BCVuniandes} & 0.10 & (0.01,0.38)\\
18 & \textit{EdwardMa12593}& 0.31 & (0.01,0.69) & 18& \textit{RegionTec}& 0.29 & (0.00,0.50)\\
19 & \textit{RegionTec} & 0.57 & (0.19,0.73) & 19& \textit{EdwardMa12593}&  0.08 & (0.01,0.18)\\

\bottomrule
\end{tabular}
\end{table}


\subsection{Impact of the challenge winner}
In the two years after the challenge, the winning algorithm, \textit{nnU-Net} (with sometimes minor modification) competed in a total of 53 further segmentation tasks. The method won 33 out of 53 tasks with a median rank of 1 (interquartile range (IQR) of (1;2)) in the 53 tasks \citep{isensee2021nnu}, for example being the winning method of the famous BraTS challenge in 2020\footnote{Team Name: MIC\_DKFZ, https://www.med.upenn.edu/cbica/brats2020/rankings.html}. This confirmed our hypothesis that a method capable of performing well on multiple tasks will generalize well to a previously unseen task and potentially outperform a custom-designed solution.
The method further became the new state-of-the-art method and was used in several segmentation challenges by other researchers. For instance, eight nnU-Net derivatives were ranked in the top 15 algorithms of the 2019 Kidney and Kidney Tumor Segmentation Challenge (KiTS)\footnote{https://kits19.grand-challenge.org/} \citep{isensee2021nnu}, the MICCAI challenge with the most participants in the year 2019. Nine out of the top ten algorithms in the COVID-19 Lung CT Lesion Segmentation Challenge 2020 (COVID-19-20)\footnote{https://covid-segmentation.grand-challenge.org/} built their solutions on top of nnU-Net (98 participants in total). As demonstrated in~\citep{ma2021cutting}, nine out of ten challenge winners in 2020 built solutions on top of nnU-Net.


\section{Discussion}
\label{sec:discussion}
We organized the first biomedical image segmentation challenge, in which algorithms competed in ten different disciplines. We showed that it is indeed possible that one single algorithm can generalize over various different applications without human-based adjustments. This was further demonstrated by monitoring the winning method for two years to show the continuation of the generalizability to other segmentation tasks.

In the following sections, we will discuss specific aspects of the MSD challenge, namely the challenge infrastructure, data set, assessment method and outcome.

\subsection{Challenge infrastructure}
The participating teams were asked to submit their results in the form of a compressed archive to the grand-challenge.org platform. For the development phase, a fully automated validation script was run for each submission and the leaderboard was updated accordingly. Each team was allowed to submit one solution per day. In contrast, for the mystery phase, only one valid submission per algorithm could be submitted to prevent over-fitting. 

Despite the above-mentioned policies, there were attempts to create multiple accounts so that a team could test their method beyond the allowed limit, a problem which was found due to result's similarity between certain accounts. Teams who were found to be evading the rules were disqualified. Identity verification and fraud detection tooling has now been added to grand-challenge.org to help organisers mitigate this problem in the future.

Possibly, a better way of controlling overfitting, or possible forms of cheating (e.g. manual labeling of results ~\citep{reinke2018exploit}) would have been to containerize the algorithms using Docker containers and for inference to be run by the organisers. This approach was unfortunately not possible at the time of the organization of MSD due to the lack of computational resources to run inference on all data for all participants. Thanks to a partnership with Amazon Web Services (AWS), the grand-challenge.org platform now offers the possibility to upload Docker container images that can participate in challenges and made available to researchers for processing new scans. With the recent announcement of a partnership between NVIDIA and the MICCAI 2020 and 2021 conferences, and the increased standardization of containers, such a solution should be adopted for further iterations of the MSD challenge.

\subsection{Challenge data set}
In the MSD, we presented a unique data set, including ten heterogeneous tasks from various body parts and regions-of-interest, numerous modalities and challenging characteristics. MSD is the largest and most comprehensive medical image segmentation data set available to date. The MSD data set has been downloaded more than 2,000 times in its first year alone, via the main challenge website.\footnote{http://medicaldecathlon.com/} The data set has recently been accepted into the AWS Open Data registry,\footnote{https://registry.opendata.aws/msd/} allowing for unlimited download and availability. The data set is also publicly available under a Creative Commons license CC-BY-SA4.0, allowing broad (including commercial) use. Due to data set heterogeneity, and usage in generalizability and domain adaptation research, it is likely to be very valuable for the biomedical image analysis community in the long term. 

Regarding limitations, the MSD data set was gathered from retrospectively acquired and labeled data from many different sources, resulting in heterogeneous imaging protocols, differences in annotation procedures, and limiting the annotations to a single human rater. While the introduction of additional annotators would have benefited the challenge by allowing inter-rater reliability estimates and possibly improve the reliability of annotations, this was not possible due to restricted resources and the scale of the data. As shown in \citep{joskowicz2019inter}, several annotators are often necessary to overcome issues related to inter-observer variability. Furthermore, the dataset only consists of radiological data, we can therefore only draw conclusions for this application. Other areas like dermatology, pathology or ophthalmology were not covered.

\subsection{Challenge assessment}
\label{subsec:dis-assessment}
Two common segmentation metrics have been used to evaluate the participant's methods, namely the DSC, an overlap measure, and the NSD, a distance-based metric. The choice of the right metrics was heavily discussed, as it is extremely important for the challenge outcome and interpretation. Some metrics are more suitable for specific clinical use-cases than others \citep{reinke2021common}. For instance, the DSC metric is a good proxy for comparing large structures but should not be used intensively for very small objects, as single-pixel differences may already lead to substantial changes in the metric scores. However, to ensure that the results are comparable across all ten tasks, a decision was taken to focus on the two above-mentioned metrics, rather than using clinically-driven task-specific metrics.

Comparability was another issue for the ranking as the number of samples varied heavily across all tasks and target ROIs, which made a statistical comparison difficult. We therefore decided to use a ranking approach similar to the evaluation of the popular BraTS challenge,\footnote{http://braintumorsegmentation.org/} which was based on a Wilcoxon-signed rank pairwise statistical test between algorithms. The rank of each algorithm was determined (independently per task and ROI) by counting the number of competing algorithms with a significantly worse performance. This strategy avoided the need of similar sample sizes for all tasks and reduced the need for task-specific weighting and score normalisation. 

Identifying an appropriate ranking scheme is a non-trivial challenge. It is important to note that each task of the MSD data set comprised one to three different target ROIs, introducing a hierarchical structure within the data set. Starting from a significance ranking for each target ROI, we considered two different aggregation schemes: (1) averaging the significance ranks across all target ROIs; (2) averaging the significance ranks per task (data set) and averaging those per-task ranks for the final rank. The drawback of (1) is that a possible bias between tasks might be introduced, as tasks with multiple target ROIs (e.g. the brain task with three target ROIs) would be over-weighted. We therefore chose ranking scheme (2) to avoid this issue. This decision was made prior to the start of the challenge, as per the challenge statistical analysis protocol. A post-challenge analysis was performed to test this decision, and results found that overall ranking structure remained unchanged. The first three ranks were preserved, only minor changes (1 to 2 ranks) were seen in a couple of examples at the middle and end of the rank list. As shown in \ref{app:dsc-subtasks}, changing the ranking scheme will typically lead to different rankings in the end, but we observed the first three ranks to be robust across various ranking variations. More complex ranking schemes were discussed among organisers, such as modeling the variations across tasks and target ROIs with a linear mixed model \citep{breslow1993approximate}. As explainability and a clear articulation of the ranking procedure was found to be important, it was ultimately decided to use significance ranking.

\subsection{Challenge outcome}
\subsubsection{Results}
The performance of algorithms varied dramatically across the different tasks, as shown in Figs.~\ref{fig:dscbox-plotsph1}, \ref{fig:dscbox-plotsph2} and \ref{app:dsc-subtasks}. For the development phase, the median algorithmic performance, defined as the median of the mean DSC, changed widely across tasks, with lowest being  the tumor mass segmentation of the pancreas data set (0.21, Table~\ref{tab:meanDSC-subtasks-pancreas-ph1}) and the highest median for the liver segmentation (0.94, Table~\ref{tab:meanDSC-subtasks-liver-ph1}). The performance drop was much more modest for the best performing method \textit{nnU-Net} (0.52 and 0.93 median DSC for the pancreas mass and liver ROI, respectively), demonstrating that methods have varying degrees of learning resiliency to the challenges posed by each task. The largest difference within one task was also obtained for the pancreas data set, with a median of the mean DSC of 0.69 for the pancreas ROI, and 0.21 for the pancreas tumor mass, which is likely explained by the very small relative intensity difference between the pancreas and its tumor mass. 
In the mystery phase, colon cancer segmentation received the lowest median DSC (0.16, Table~\ref{tab:meanDSC-subtasks-colon-ph2}), and the spleen segmentation the highest median DSC (0.94, Table~\ref{tab:meanDSC-subtasks-spleen-ph2}). Similarly to the development phase, a much smaller drop in performance (0.56 and 0.96 for colon and spleen respectively) was observed in the top ranking method. Most of the observed task-specific performances reflect the natural difficulty and expected inter-rater variability of the tasks: Liver and spleen are large organs that are easy to detect and outline \citep{campadelli2009liver}, whereas pancreas and colon cancers are much harder to segment as annotation experts themselves often do not agree on the correct outlines \citep{sirinukunwattana2017gland, re2011enhancing}. We also observed that the challenging characteristics of each task (presented in Fig.~\ref{fig:overview}) had some non-trivial effect on algorithmic performance, a problem which was exacerbated in lower-ranking methods. For example, some methods struggled to segment regions such as the lung cancer mass, pancreas mass, and colon cancer primaries, achieving a mean DSC below 0.1. These regions, characterized by small, non-obvious and heterogeneous masses, appear to represent a particularly challenging axis of algorithmic learning complexity. The number of subjects in the training dataset (only 30 subjects for the heart task), the size and resolution of the images (large liver images and small hippocampus images), and complex region shapes (e.g. brain tumours) were not found to introduce significant inter-team performance differences. 

\subsubsection{Methods}
As summarised in Fig.~\ref{fig:stackedplots}, \textit{nnU-Net} was ranked first on both the development and mystery phases. Under the proposed definition of a \textit{"generalizable learner"}, the winning method was found to be the most generalizable approach across all MSD tasks given the comparison methodology, with a significant performance margin. The \textit{K.A.V.athlon} and \textit{NVDLMED} teams were ranked second and third during the development phase, respectively; their ranks were swapped (third and second, respectively) during the mystery phase. We observed small changes in team rankings between the development and mystery phases for top ranking teams; within the top 8 teams, no team changed their ranking by more than 2 positions from the development to the  mystery phase. This correlation between development and mystery rankings suggest limited amount of methodological overfitting to the development phase, and that the proposed ranking approach is a good surrogate of expected  task performance. 
We observed some algorithmic commonalities between top methods, such the use of ensembles, intensity and spatial normalization augmentation, the use of Dice loss, the use of Adam as an optimiser, and some degree of post-processing (e.g. region removal). While none of these findings are surprising, they provide evidence towards a reasonable choice of initial parameters for new methodological developments. We also observed that the most commonly applied architecture across participants was the U-Net, used by 64\% of teams. Some evidence was found that architectural adjustments to the baseline U-Net approach are less important than other relevant algorithmic design decisions, such as data augmentation and data set split/cross-validation methodology, as demonstrated by the winning methodology. Note that similar findings, albeit in a different context and applied to ResNet, have been recently observed \citep{bello2021revisiting}. 

\subsection{The years after the challenge}
Following the challenge event at MICCAI 2018, the competition was opened again for rolling submissions. This time participants were asked to submit results for all ten data sets\footnote{https://decathlon-10.grand-challenge.org/} in a single phase. In total, 742 users signed up. To restrict the exploitation of the submission system for other purposes, only submissions with per-task metric values different from zero were accepted as valid, resulting in only 17 complete and valid submissions. In order to avoid overfit but still allow for some degree of methodological development, each team was allowed submit their results 15 times. The winner of the 2018 MSD challenge (\textit{nnU-Net}, denoted as \textit{Isensee} on the live challenge), submitted to the live challenge leaderboard on the 6$^{\text{th}}$ of December 2019, and held the first position for almost one year, until the 30$^{\text{th}}$ of October 2020.     

Since for the live challenge teams were allowed to tune their method on all ten data sets, the minimum value of the data set specific median DSC improved quite substantially from the 2018 MSD challenge, as shown in Fig \ref{fig:box-plot2018}. The two hardest tasks during the 2018 MSD challenge were the segmentation of the tumor inside the pancreas, with an overall median of the mean DSC of 0.21 over all participants (0.37 for the top five teams) and the segmentation of the colon cancer primaries, with an overall median of the mean DSC of 0.16 over all participants (0.41 for the top five teams). The worst task for the rolling challenge was the segmentation of the non-enhancing tumor segmentation inside the brain, with a median DSC of 0.47.

At the other end of the spectrum was the spleen segmentation task, where the median task DSC over all participants was 0.94 during the 2018 challenge, and improved to 0.97 for the rolling challenge. These observations suggest that the ability for multiple methods to solve the task has improved, with methods performing slightly better on harder tasks and significantly better on easy tasks. 

In 2019 and 2020, the rolling challenges have resulted in three methods that superseded the winning results of the 2018 MSD challenge. Within these two follow-up years, two main trends were observed: the first major trend is the continuous and gradual improvement of "well performing" algorithms, such as the heuristics and task fingerprinting of the \textit{nnU-Net} method; the second major trend that was observed was the rise of Neural Architecture Search (NAS) \citep{elsken2019neural} among the top teams. More specifically, both the third and the current \citep{yufanHE} (as of April 2021) leader of the rolling challenge used this approach. NAS optimizes the network architecture itself to each task in a fully automated manner. Such an approach uses a network-configuration fitness function that is optimised independently for each task, thus providing an empirical approach for network architectural optimisation. When compared to heuristic methods (e.g. \textit{nnU-Net}), NAS appears to result in improved algorithmic performance at the expense of increased computational cost. 

\section{Conclusion}
Machine-learning based semantic segmentation algorithms are becoming increasingly general purpose and accurate, but have historically required significant field-specific expertise to use. The MSD challenge was set up to investigate how accurate fully-automated image segmentation learning methods can be on a plethora of tasks with different types of task complexity. Results from the MSD challenge have demonstrated that fully automated methods can now achieve state-of-the-art performance without the need for manual parameter optimisation, even when applied to previously unseen tasks. A central hypothesis of the MSD challenge---that an algorithm which works well and automatically on several tasks should also work well on other unseen tasks---has been validated among the challenge participants and across tasks. This hypothesis was further corroborated by monitoring the generalizability of the winning method in the two years following the challenge, where we found that \textit{nnU-Net} achieved state-of-the-art performance on many tasks including against task-optimized networks. While it is important to note that many classic semantic segmentation problems (e.g. domain shift and label accuracy) remain, and that methodological progress (e.g. NAS and better heuristics) will continue pushing the boundaries of algorithmic performance and generalizability, the MSD challenge has demonstrated that the training of accurate semantic segmentation networks can now be fully automated. This commoditization of semantic segmentation methods has the potential to allow non machine learning experts (e.g. clinicians, medical physicists,and applied scientists) to better contribute to, and possibly even independently develop, these techniques.

\section{Acknowledgements}
Part of this work was funded by the Helmholtz Imaging Platform (HIP), a platform of the Helmholtz Incubator on Information and Data Science.
We would like to thank Minu D. Tizabi for proof-reading the manuscript.
This research was supported by the Bavarian State Ministry of Science and the Arts and coordinated by the Bavarian Research Institute for Digital Transformation.
Team CerebriuDIKU gratefully acknowledges support from the Independent Research Fund Denmark through the project U-Sleep (project number 9131-00099B).
Ronald M. Summers is supported by the Intramural Research Program of the National Institutes of Health Clinical Center
Research reported in this publication was partly supported by the National Institutes of Health (NIH) under award numbers NCI:U01CA242871, NCI:U24CA189523, NINDS:R01NS042645. The content of this publication is solely the responsibility of the authors and does not represent the official views of the NIH.
The method presented by BCVUniandes was made in collaboration with Silvana Castillo, from Universidad de los Andes.
James Meakin received grant funding from Amazon Web Services.

\section*{Author information}
These authors contributed equally to this work (shared first authors): Michela Antonelli, Annika Reinke
These authors contributed equally to this work (shared senior authors): Lena Maier-Hein, M. Jorge Cardoso

Corresponding author: Michela Antonelli (email address: \texttt{michela.antonelli@kcl.ac.uk}).

\appendix

\section{Challenge organization}
\label{app:organization}
The MSD challenge was organized in the scope of MICCAI 2018, held in Granada, Spain. It was organized by M. Jorge Cardoso (King's College London), Amber Simpson (Memorial Sloan Kettering Cancer Center), Olaf Ronneberger (Google Deep mind), Bjoern Menze (Technische Universität München), Bram van Ginneken (Radboud University Medical Center), Bennett Landman (Vanderbilt University), Geert Litjens (Radboud University Medical Center), Keyvan Farahani (National Institutes of Health), Ronald M. Summers (National Institutes of Health Clinical Center), Lena Maier-Hein (DKFZ German Cancer Research Center), Annette Kopp-Schneider (DKFZ German Cancer Research Center), Spyridon Bakas (CBICA, University of Pennsylvania) and Michela Antonelli (King's College London).

The challenge was organized as an open call event, i.e. after the challenge event at MICCAI, the challenge still accepted and evaluated submissions. For the submission itself, the grand-challenge.org platform was used,\footnote{https://decathlon-10.grand-challenge.org} whereas all other information was given on a separated website.\footnote{http://medicaldecathlon.com}

The participation policies of the MSD allowed only fully automatic methods without task-specific manual parameter settings to submit. In addition, there was no restriction in using other data sources to pre-train the individual methods, as long as that data was not modified per task. Only one team was allowed per research lab, as to avoid bypassing submission count restrictions. The first runner method of each phase, and the runner up of the mystery phase all received an NVIDIA Titan V prize. Ranking and results on each data set for each method were announced publicly at the challenge event at MICCAI and the post-challenge leaderboard is also publicly available. Finally, we asked the participating teams to fill out a form with details about their methods. All team members replying to the survey were listed as co-authors of the paper. Participants were allowed to publish their methods independently from the challenge paper.

All teams were asked to submit the results of the development phase as a compressed archive to the \url{grand-challenge.org} platform. A fully automated validation script was run for each submission immediately after submission and results were published on the development phase leaderboard. Each team was allowed one submission per day to partially mitigate overfitting. The last submission of each team by the development phase deadline was used for validation. Teams were then asked to submit details of their methods prior to being given access to the mystery phase data. The submission deadline for the mystery phase results was set to two weeks after the mystery phase data release, also in  compressed archive form. Only a single valid submission was accepted for each the mystery phase participant, and results for the mystery phase were only revealed at the public challenge event during MICCAI 2018, in Granada, Spain. 

The implementation of the metrics used in the challenge, namely the DSC and NSD, were provided as a Python Notebook \citep{MSDmetrics} by the challenge organisers, prior to the challenge deadlines. The algorithms for statistical validation were also provided.

As some of the participating teams were working under intellectual property restrictions, it was decided that public code availability was not mandatory for participation as to maximise participation; participants were, however, encouraged to make their code available to the public.  

The challenge was organised without specific funding, and mostly via in-kind time contributions of its organisers. The challenge was sponsored by NVIDIA, who provided the graphics processing unit (GPU) cards as awards (approximate value of \$7,500), Google DeepMind, who provided an in-kind implementation of the NSD metric, and by RSIP Vision, who provided media support and challenge dissemination. None of the sponsors had any influence in the organisation of the challenge, nor were they given any form of privileged access to either the data or any other type of information. 

Only two of the organisers, both from KCL, had access to all test cases; namely M. Jorge Cardoso and Michela Antonelli. The KCL organisers committed to not participate in the challenge. Only two copies of the full test data currently exist (beyond the data providers of each independent task), one at KCL's servers as a backup, and one on the \url{grand-challenge.org} validation server.

\section{Method details}
\label{app:methods}
In the following, we provide details for the remaining teams that submitted a description of their methods. Note that the methods from the top three teams are presented in Section~\ref{subsec:methoddescription} of the main paper. 

\subsubsection{AI-MED}
The team used the QuickNAT with added Conditional Random Fields (CRF) \citep{QuickNat}.
The DSC loss was combined with the cross entropy loss and the SGD optimizer was used. No data augmentation or ensembling techniques were employed. 

\subsubsection{BCVuniandes}
The team employed DeepMedic \citep{kamnitsas2016deepmedic} as base architecture by using two identical parallel pathways with multi-scale analysis. Each pathway had four stages and all the intermediate outputs were resized and concatenated to be processed by two fully connected layers. The team did not apply augmentation or ensembling techniques, but used a softmax cross-entropy loss and the Adam optimizer.

\subsubsection{CerebriuDIKU}
The method used the standard 2D U-Net architecture with added batch normalization layers intervening each double-convolution- and up-convolution block and nearest neighbour up-sampling~\citep{perslev2019one}.
Two augmentations were applied, namely non-linear deformation and multi-planar sampling of 2D image planes. Furthermore, the cross entropy loss and the Adam optimizer were applied. 
Averaging multiple runs of the same architecture was the used ensembling strategy.
The key aspect of the method was the multi-planar training that allowed for a huge number of anatomically relevant images to be augmented during training, exposing the model to a broader representation of the 3D image volume while maintaining the parameter (and computational) efficiency of 2D kernels. Optimizing over multiple planes increased the complexity of the target function making overfitting less likely but maintained performance by doing so through the exposure of the model to additional data. The model learned to segment the target as seen from multiple views and can therefore be used to predict the same target multiple times.

\subsubsection{Lesswire1}
The team used a U-Net architecture with concatenated lower resolution features before upscaling and the transpose convolution for upscaling. 
Test-time augmentation was applied and a depth-wise cross-entropy was the loss function which consisted of a weighted sum of depth-wise cross entropy and L2 norm. Adam was used and no ensembling techniques were applied.
The key idea of the model was that it was designed to operate on any volume regardless of the number of slices. No assumption was made to limit the model to a specific organ. Losses associated with anomalies were weighted based on the training data adaptively so that the model loss was not specific to a certain organ.

\subsubsection{LfB}
The team modified the U-Net architecture with residual connections per block, deep supervision (multi-level generation of segmentations), and instance normalization over batch normalization \citep{lfb}. 
The following augmentation strategies were used: affine transformation, non-linear deformation, noise addition, rotation, and random crop; moreover, if anisotropy, elastic deformation field was scaled to yield isotropic world coordinate scaling. DSC loss and Adam were used. No ensembling strategy was applied. 
The key point of the method was to do as little task-specific engineering as possible (i.e. during training only resampled to median voxel spacing, got median shape, and inferred model geometry/architecture from median shape; for inference, resampled to median voxel spacing, applied patch-wise prediction, and resampled to original voxel spacing).

\subsubsection{LS Wang's}
The team proposed a modification of the U-Net, more specifically a Nested Dilation Network (NDN) which was applied to multiple segmentation tasks and multiple modalities. The Residual Blocks Nested were designed with dilations (RnD Blocks) which catch larger receptive field in the first few layers to boost shallow semantic information~\citep{NDN}. 
The following augmentation strategies were applied: affine transformations, non-linear deformation, noise addition, histogram transformation, geometric left-right flip, and random crop. 
The team used a modification of focal loss and the Adam optimizer and applied averaging multiple runs of the same architecture as ensambling strategy. 
The unique configuration was the key to cope with the ten different tasks.

\subsubsection{Lupin}
To take advantage of the ability of the U-Net in combining high-level features and low-level features, the team added deep-supervision during training on decoding path. 
Affine transformation, noise addition, and random crop were applied as augmentation techniques. Focal loss and Adam were used. No ensembling strategy was employed. 

\subsubsection{MIMI}
U-Net was the base network architecture of this team, which was used with the following additional components: 1) residual Block: skip connections inside convolutional blocks; 2) auxiliary losses for deep supervision; 3) leaky ReLU; 4) dropout inside decoders. The team applied affine transformations, geometric left-right flip, and random crop as augmentation strategy. DSC loss and the Adam optimizer were employed. They average multiple runs of the same architecture as ensambling strategy.
Using Multi-Task Learning (MTL) was the key point of this approach. Representations among related tasks were shared and a better generalization achieved. 

\subsubsection {SIAT\_MIDS}
A modified version of V-Net was used. In particular, they used a two-level V-Net model with batch normalization after each convolution in the second layer and set the number of feature maps as 32 after the first convolution. The input size was still 128×128×64 except for those tasks who have ROI larger than 128 in x- or y-axis, otherwise a size of 160×160×64 was used for the memory limitation.
Two augmentation techniques were applied, namely non-linear deformation and histogram transformation.
DSC loss and weighted softmax were used. SGD was the optimizer. No ensambling techniques were applied.
The key aspect of the method was using a two-stage coarse-to-fine method that is a general approach for medical volume segmentation. 
They proposed an automatic ROI extraction technique based on the initial coarse segmentation in the first level. The input size of the V-Net in the second level had two types which were decided according to the extracted ROI. 

\subsubsection{VST}
The team used different models depending on the task. More specifically, they used a 2.5D U-Net, a 3D CNN and a 3D U-Net for detection, classification, and  segmentation, respectively. 
2.5D U-Net considered a total of 5 slices adjacent in an axial direction and output the segmentation probability at the central position among the 5 slices. 3D CNN considered 3D voxels and classified whether the target objects included the voxel or not.
Affine transformations (rotation, translation, scale) were applied as augmentation techniques. The loss function was different for each model: weighted cross entropy for the 2.5D U-Net, cross entropy for 3D CNN, and weighted cross entropy for 3D U-Net. The optimizer was SGD for 3D CNN and Adam for the other models. Multiple-architecture weighted averaging was used as ensebling technique. 

\subsubsection{Whale}
An ensemble of both 3D and 2D U-Nets was used, in which the 3D U-Net was shallow while the 2D U-Net was much deeper. Geometric left-right flip and random crop were applied as augmentation techniques. The team employed cross entropy loss with larger weights to minority classes and the SGD optimizer.
Multiple-architecture non-weighted averaging was the ensembling strategy.
The method relied on the use of U-Net which is very general and its effectiveness has been demonstrated on many tasks.

\newpage
\section{DSC values and rankings for all target regions}
\label{app:dsc-subtasks}
\begin{table}[H]
\scriptsize
\centering
\caption{Mean Dice Similarity Coefficient (DSC) values for all participating teams for all tasks (edema, non-enhancing tumor, and enhancing tumor) of the \textbf{brain} data set (the development phase).}
\label{tab:meanDSC-subtasks-brain-ph1}
\begin{center}
\begin{tabular}{lccc}
\toprule
\textbf{Algorithm} & \textbf{Edema} & \textbf{Non-enhancing tumor} & \textbf{Enhancing tumor} \\
\midrule
\textit{nnU-Net} & 0.68 & 0.48 & 0.68 \\
\textit{NVDLMED}& 0.68 & 0.45 & 0.68 \\
\textit{K.A.V.athlon}& 0.66 & 0.47 & 0.67 \\  
\textit{LS Wang's Group} & 0.68 & 0.46 & 0.66 \\
\textit{MIMI}& 0.65 & 0.44 & 0.66 \\
\textit{CerebriuDIKU}& 0.69 & 0.43 & 0.67 \\
\textit{Whale} & 0.64 & 0.29 & 0.23 \\
\textit{UBIlearn}&0.65 & 0.37 & 0.62 \\
\textit{Lupin}& 0.66 & 0.42 & 0.64 \\
 \textit{Jiafucang} &  0.33 & 0.27 & 0.31 \\
\textit{LfB}& 0.59 & 0.44 & 0.62\\
\textit{A-REUMI01}& 0.64 & 0.40 & 0.64 \\
\textit{VST}& 0.54 & 0.29 & 0.63 \\
\textit{AI-Med}& 0.64 & 0.35 & 0.61 \\
\textit{Lesswire1} & 0.63 & 0.41 & 0.58 \\
\textit{BUT}& 0.64 & 0.35 & 0.62 \\
\textit{RegionTec}& 0.42 & 0.35 & 0.55 \\
\textit{BCVuniandes} & 0.69 & 0.43 & 0.65 \\
\textit{EdwardMa12593}& 0.37 & 0.01 & 0.18 \\
\midrule
\textbf{Median} & \textbf{0.64} & \textbf{0.41} & \textbf{0.63} \\
\bottomrule
\end{tabular}
\end{center}
\end{table} 
\newpage
\begin{figure}[H]
\hspace{-2cm}
    \centering
    \begin{minipage}[t]{0.55\textwidth}
    \includegraphics[width=1.2\textwidth]{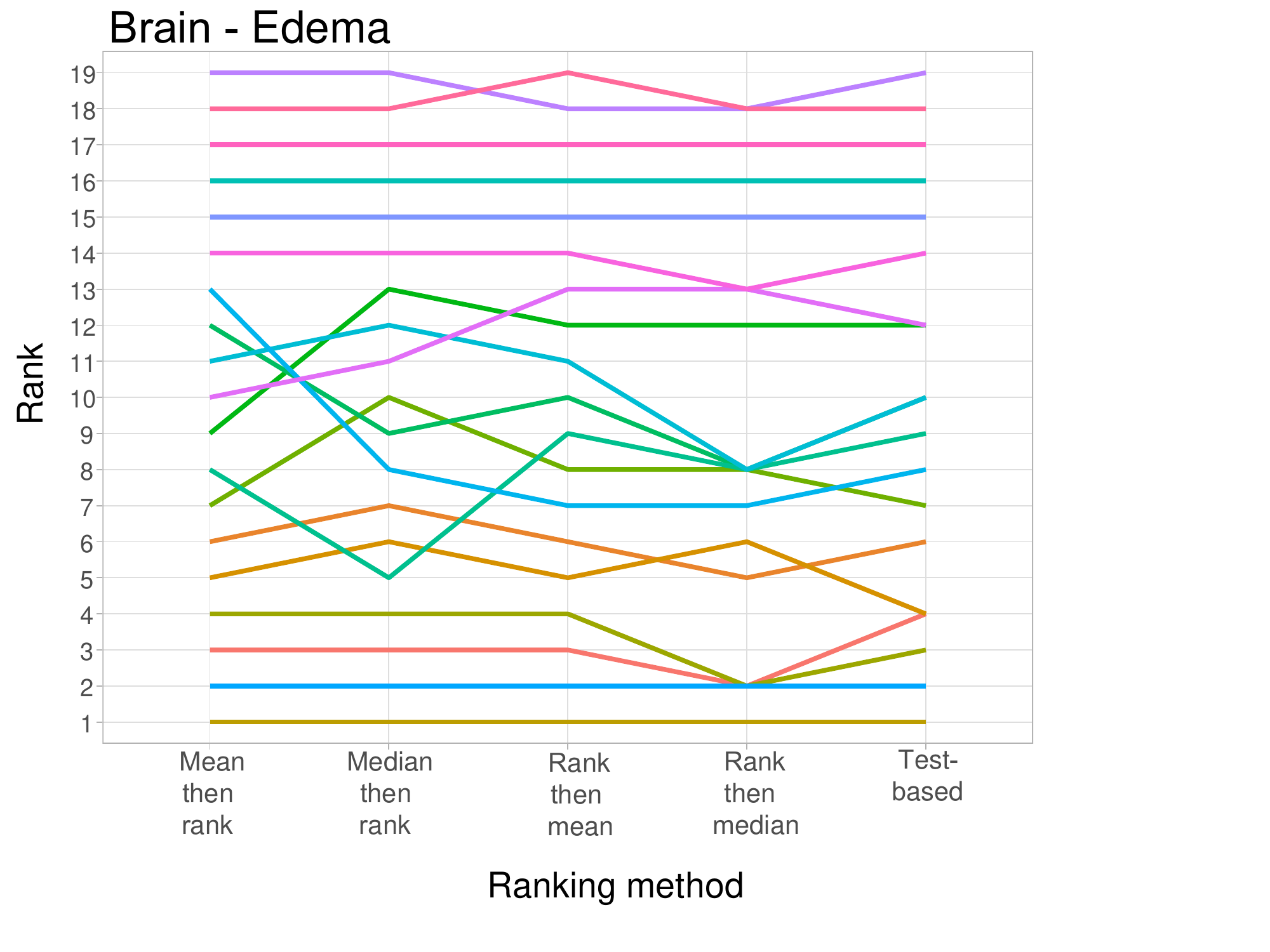}
    \end{minipage}
    \begin{minipage}[t]{0.55\textwidth}
    \includegraphics[width=1.2\textwidth]{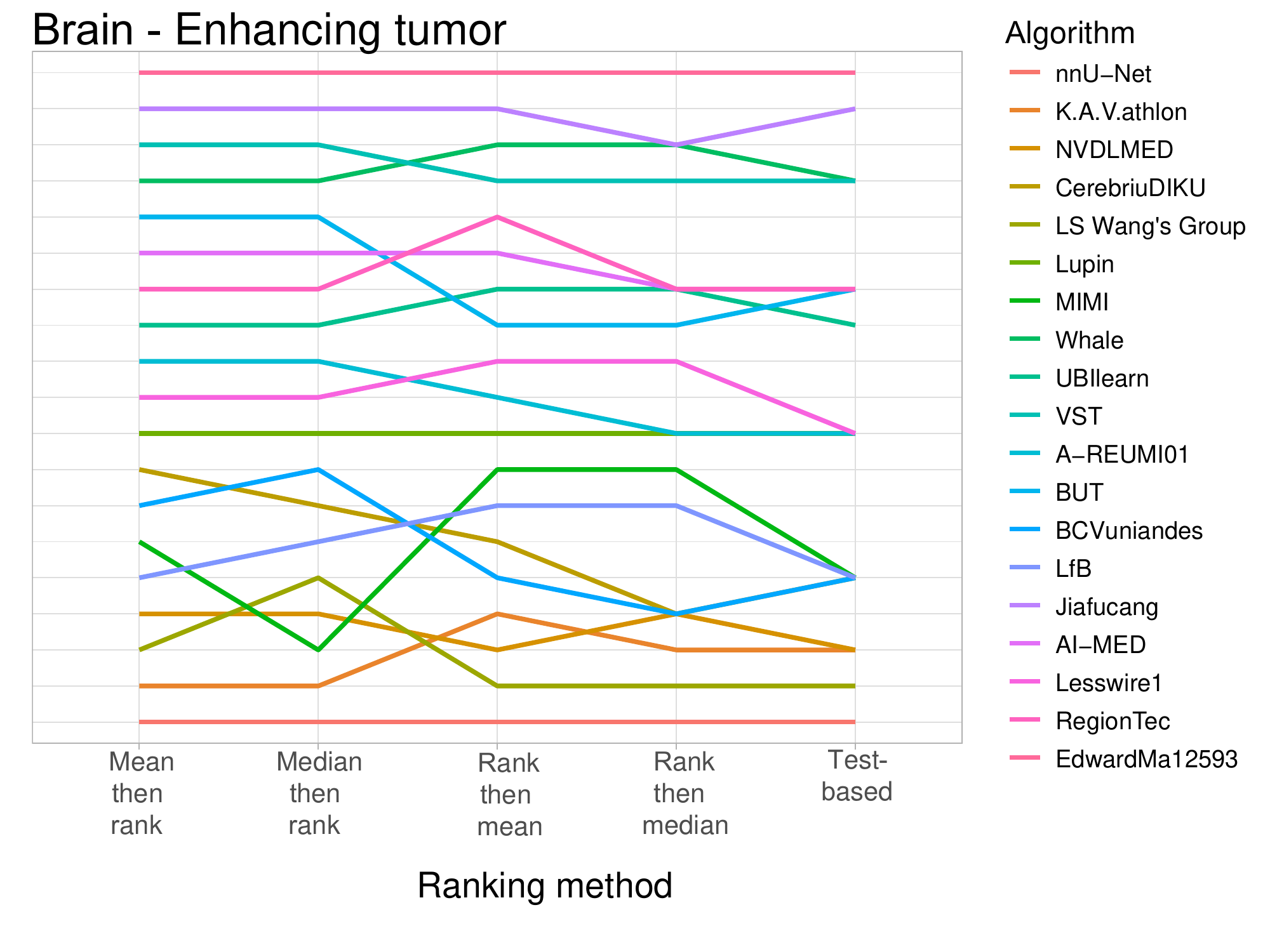}
    \end{minipage}
    \begin{minipage}[t]{0.55\textwidth}
    \includegraphics[width=1.2\textwidth]{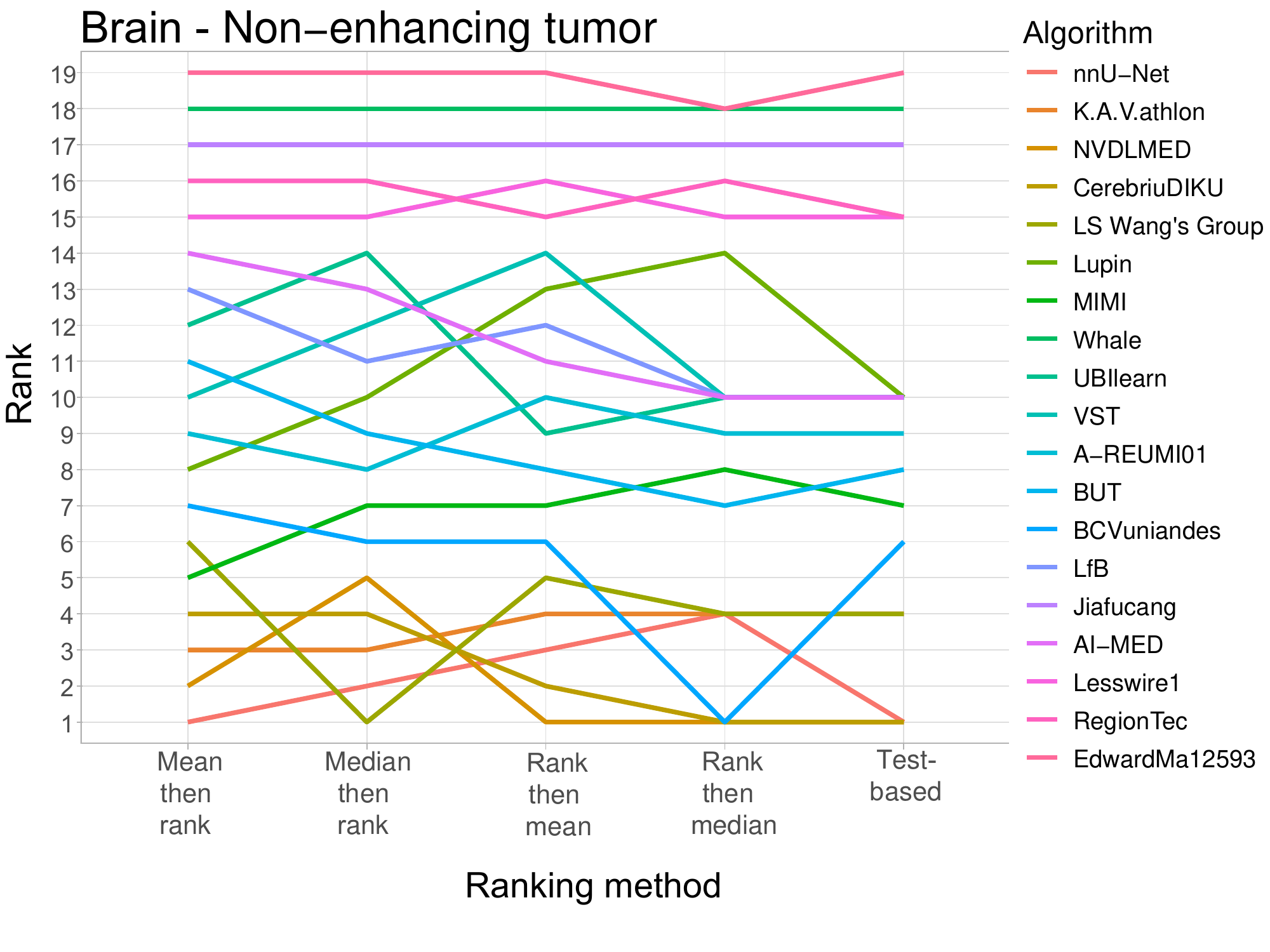}
    \end{minipage}
    \caption{Line plots visualizing rankings robustness across different ranking methods for the \textbf{brain} task. Each algorithm is represented by one colored line. For each ranking method encoded on the x-axis, the height of the line represents the corresponding rank. Horizontal lines indicate identical ranks for all methods under all ranking criteria.}
    \label{fig:line-plots-brain}
\end{figure}

\begin{table}[H]
\centering
\scriptsize
\caption{Mean Dice Similiarty Coefficient (DSC) values for all participating teams for all tasks (left atrium) of the \textbf{heart} data set (the development phase).}
\label{tab:meanDSC-subtasks-cardiac-ph1}
\begin{center}
\begin{tabular}{lc}
\toprule
\textbf{Algorithm} & \textbf{Left atrium} \\
\midrule
\textit{nnU-Net} & 0.93 \\
\textit{NVDLMED}& 0.92  \\
\textit{K.A.V.athlon}& 0.92 \\  
\textit{LS Wang's Group} & 0.89\\
\textit{MIMI} & 0.89\\
\textit{CerebriuDIKU} & 0.89\\
\textit{Whale} & 0.89 \\
\textit{UBIlearn} & 0.91 \\
\textit{Lupin} & 0.92 \\
 \textit{Jiafucang} & 0.88\\
\textit{LfB} & 0.91\\
\textit{A-REUMI01} & 0.89 \\
\textit{VST} & 0.89\\
\textit{AI-Med} & 0.84\\
\textit{Lesswire1} & 0.72\\
\textit{BUT} & 0.76 \\
\textit{RegionTec} & 0.63\\
\textit{BCVuniandes} & 0.80 \\
\textit{EdwardMa12593} & 0.73 \\
\midrule
\textbf{Median} & \textbf{0.89}\\
\bottomrule
\end{tabular}
\end{center}
\end{table} 

\begin{figure}[H]
    \centering
    \includegraphics[width=0.6\textwidth]{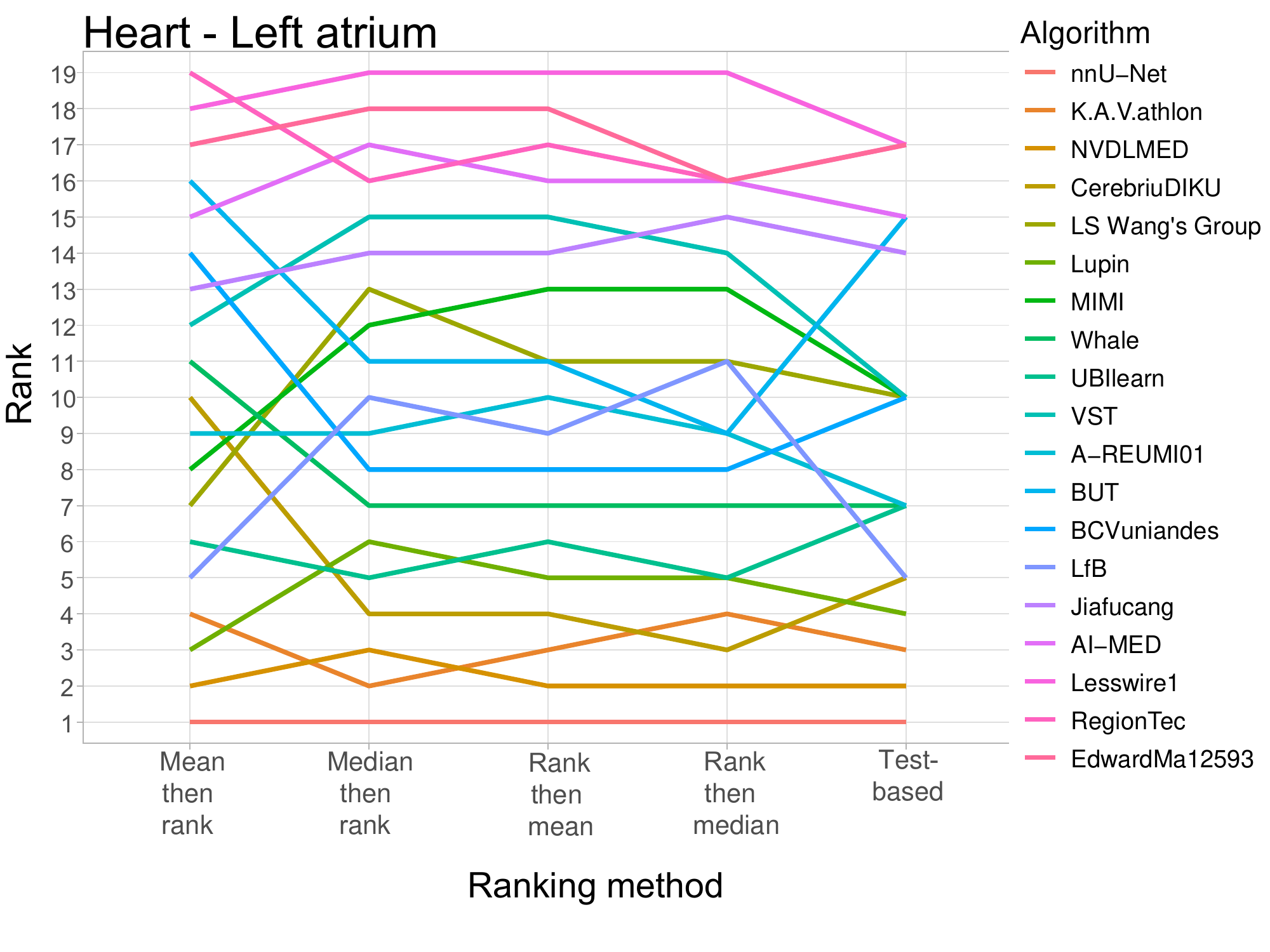}
    \caption{Line plots visualizing rankings robustness across different ranking methods for the \textbf{heart} task. Each algorithm is represented by one colored line. For each ranking method encoded on the x-axis, the height of the line represents the corresponding rank. Horizontal lines indicate identical ranks for all methods.}
    \label{fig:line-plots-cardiac}
\end{figure}

\begin{table}[H]
\scriptsize
\centering
\caption{Mean Dice Similiarty Coefficient (DSC) values for all participating teams for all tasks (anterior, posterior) of the \textbf{hippocampus} data set (the development phase).}
\label{tab:meanDSC-subtasks-hippo-ph1}
\begin{center}
\begin{tabular}{lcc}
\toprule
\textbf{Algorithm} & \textbf{Anterior} & \textbf{Posterior} \\
\midrule
\textit{nnU-Net} & 0.90 & 0.89 \\
\textit{NVDLMED}& 0.88 & 0.87 \\
\textit{K.A.V.athlon}& 0.89 & 0.89\\  
\textit{LS Wang's Group} & 0.90 & 0.89\\
\textit{MIMI}& 0.89 & 0.88\\
\textit{CerebriuDIKU}& 0.89 & 0.88 \\
\textit{Whale} & 0.86 & 0.85 \\
\textit{UBIlearn}& 0.89 & 0.85\\
\textit{Lupin}& 0.89 & 0.88 \\
 \textit{Jiafucang} & 0.88 & 0.87 \\
\textit{LfB}& 0.79 & 0.84\\
\textit{A-REUMI01}& 0.88 & 0.86 \\
\textit{VST}& 0.86 & 0.87 \\
\textit{AI-Med}& 0.84 & 0.73 \\
\textit{Lesswire1} & 0.80 & 0.77 \\
\textit{BUT}& 0.83 & 0.83 \\
\textit{RegionTec}&  0.71 & 0.70 \\
\textit{BCVuniandes} & 0.89 & 0.88 \\
\textit{EdwardMa12593} & 0.87 & 0.87 \\
\midrule
\textbf{Median} & \textbf{0.88} & \textbf{0.89}  \\
\bottomrule
\end{tabular}
\end{center}
\end{table} 

\begin{figure}[H]
\hspace{-2cm}
    \centering
    \begin{minipage}[t]{0.55\textwidth}
    \includegraphics[width=1.2\textwidth]{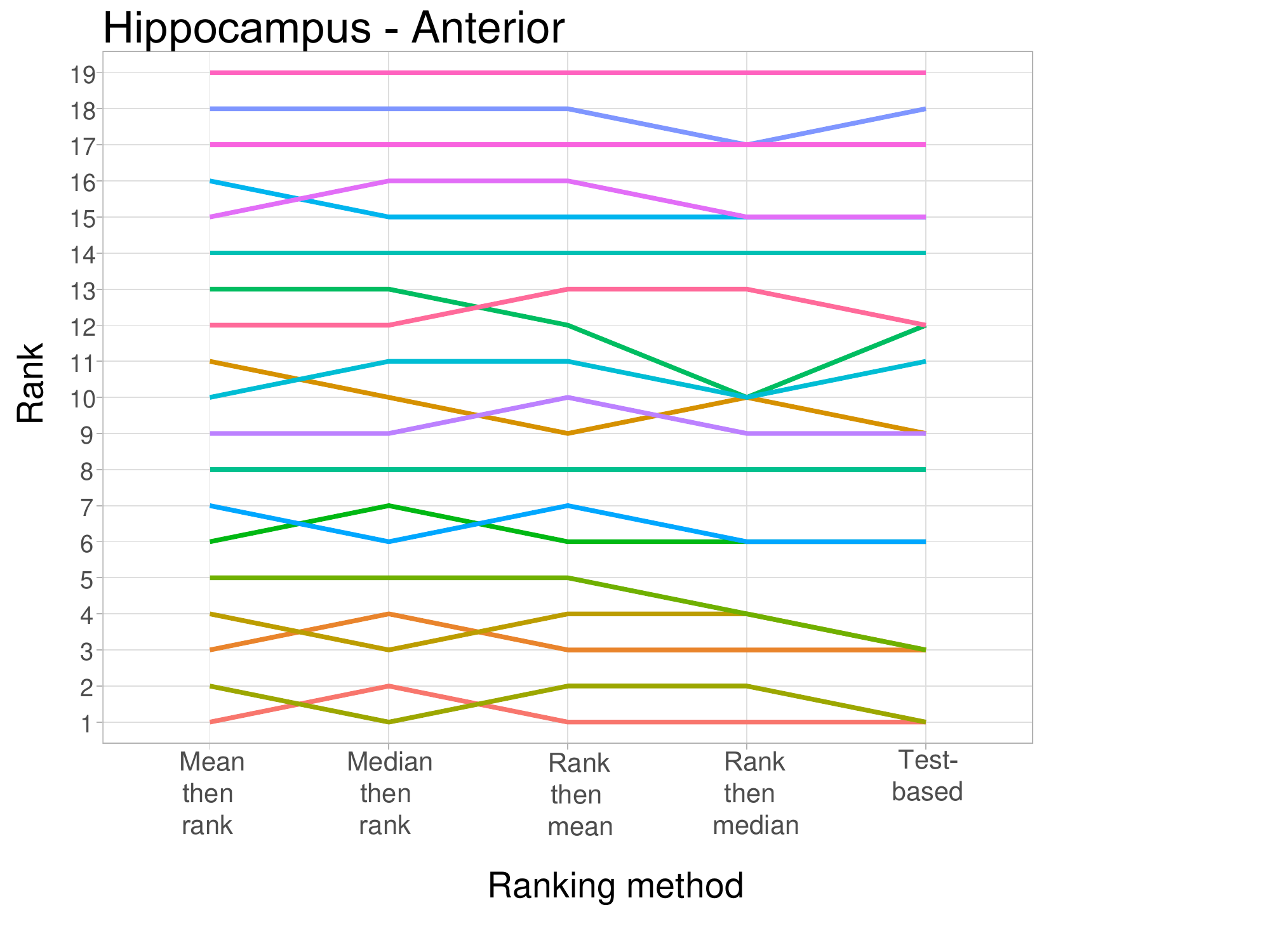}
    \end{minipage}
    \begin{minipage}[t]{0.55\textwidth}
    \includegraphics[width=1.2\textwidth]{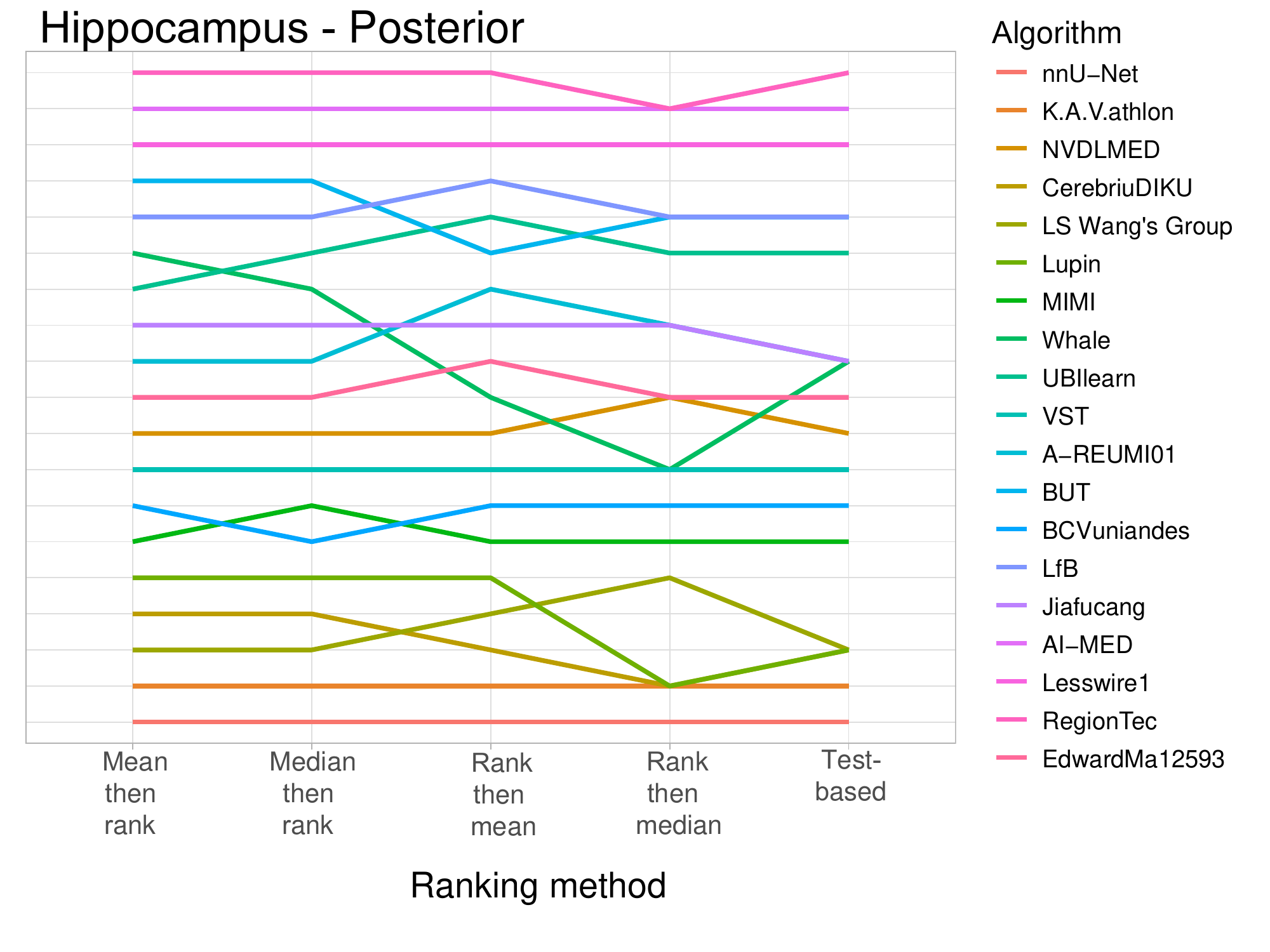}
    \end{minipage}
    \caption{Line plots visualizing rankings robustness across different ranking methods for the \textbf{hippocampus} task. Each algorithm is represented by one colored line. For each ranking method encoded on the x-axis, the height of the line represents the corresponding rank. Horizontal lines indicate identical ranks for all methods.}
    \label{fig:line-plots-hippocampus}
\end{figure}

\begin{table}[H]
\scriptsize
\centering
\caption{Mean Dice Similiarty Coefficient (DSC) values for all participating teams for all tasks (liver, cancer) of the \textbf{liver} data set (the development phase).}
\label{tab:meanDSC-subtasks-liver-ph1}
\begin{center}
\begin{tabular}{lcc}
\toprule
\textbf{Algorithm} & \textbf{Liver} & \textbf{Cancer} \\
\midrule
\textit{nnU-Net} & 0.93 & 0.74 \\
\textit{NVDLMED}& 0.95 & 0.71 \\
\textit{K.A.V.athlon}&0.94 & 0.61 \\  
\textit{LS Wang's Group} & 0.94 & 0.55\\
\textit{MIMI} & 0.94 & 0.60 \\
\textit{CerebriuDIKU}& 0.94 & 0.57 \\
\textit{Whale} & 0.94 & 0.62 \\
\textit{UBIlearn} & 0.94 & 0.50 \\
\textit{Lupin}& 0.95 & 0.61 \\
 \textit{Jiafucang} & 0.29 & 0.55 \\
\textit{LfB}& 0.90 & 0.46 \\
\textit{A-REUMI01} & 0.93 & 0.43\\
\textit{VST} & 0.93 & 0.54 \\
\textit{AI-Med} & 0.91 & 0.51 \\
\textit{Lesswire1} & 0.85 & 0.48\\
\textit{BUT}& 0.94 & 0.60 \\
\textit{RegionTec} & 0.91 & 0.32 \\
\textit{BCVuniandes} & 0.44 & 0.42\\
\textit{EdwardMa12593}& 0.81 & 0.15 \\
\midrule
\textbf{Median} & \textbf{0.94} & \textbf{0.54} \\
\bottomrule
\end{tabular}
\end{center}
\end{table} 

\begin{figure}[H]
\hspace{-2cm}
    \centering
    \begin{minipage}[t]{0.55\textwidth}
    \includegraphics[width=1.2\textwidth]{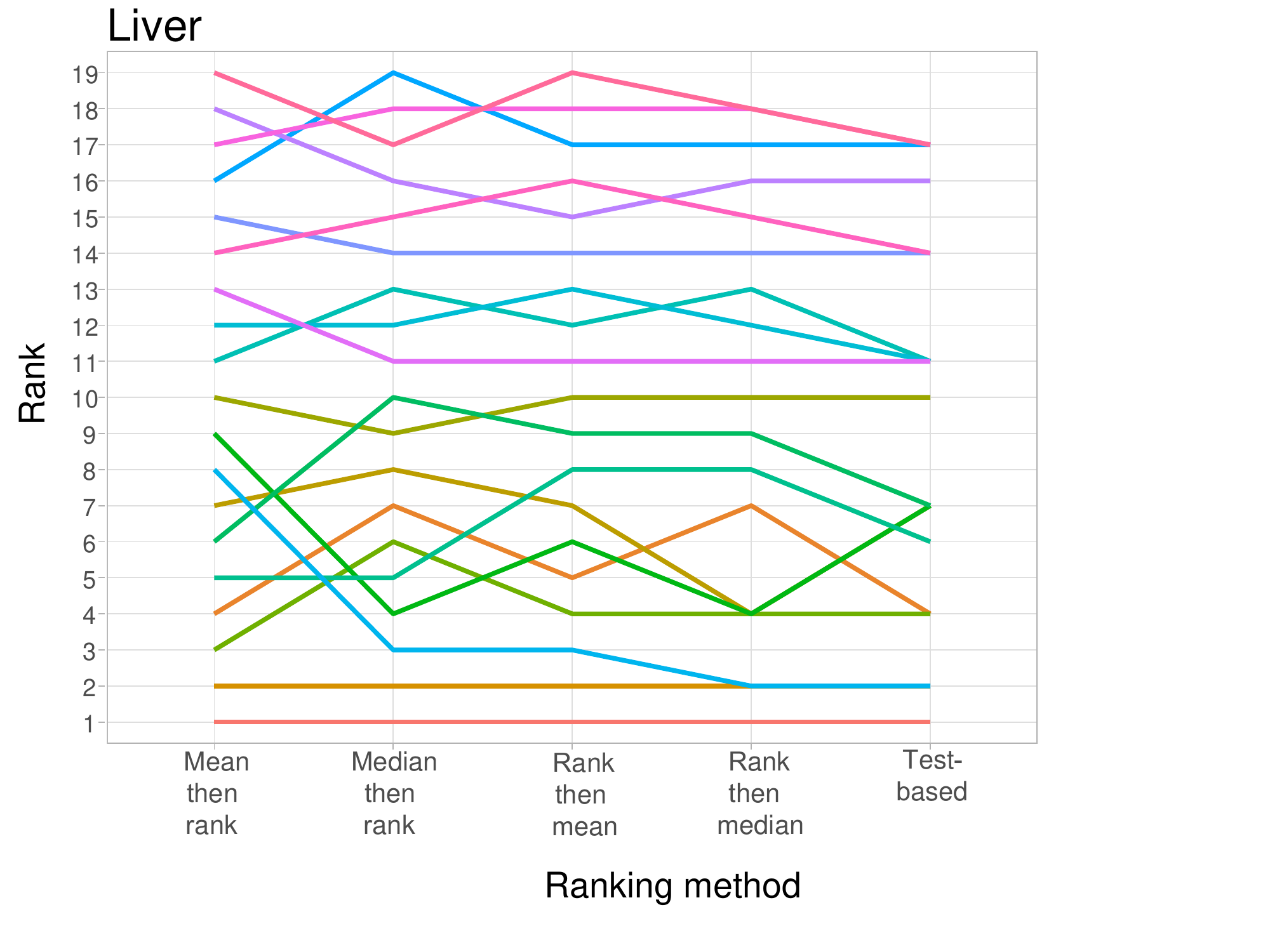}
    \end{minipage}
    \begin{minipage}[t]{0.55\textwidth}
    \includegraphics[width=1.2\textwidth]{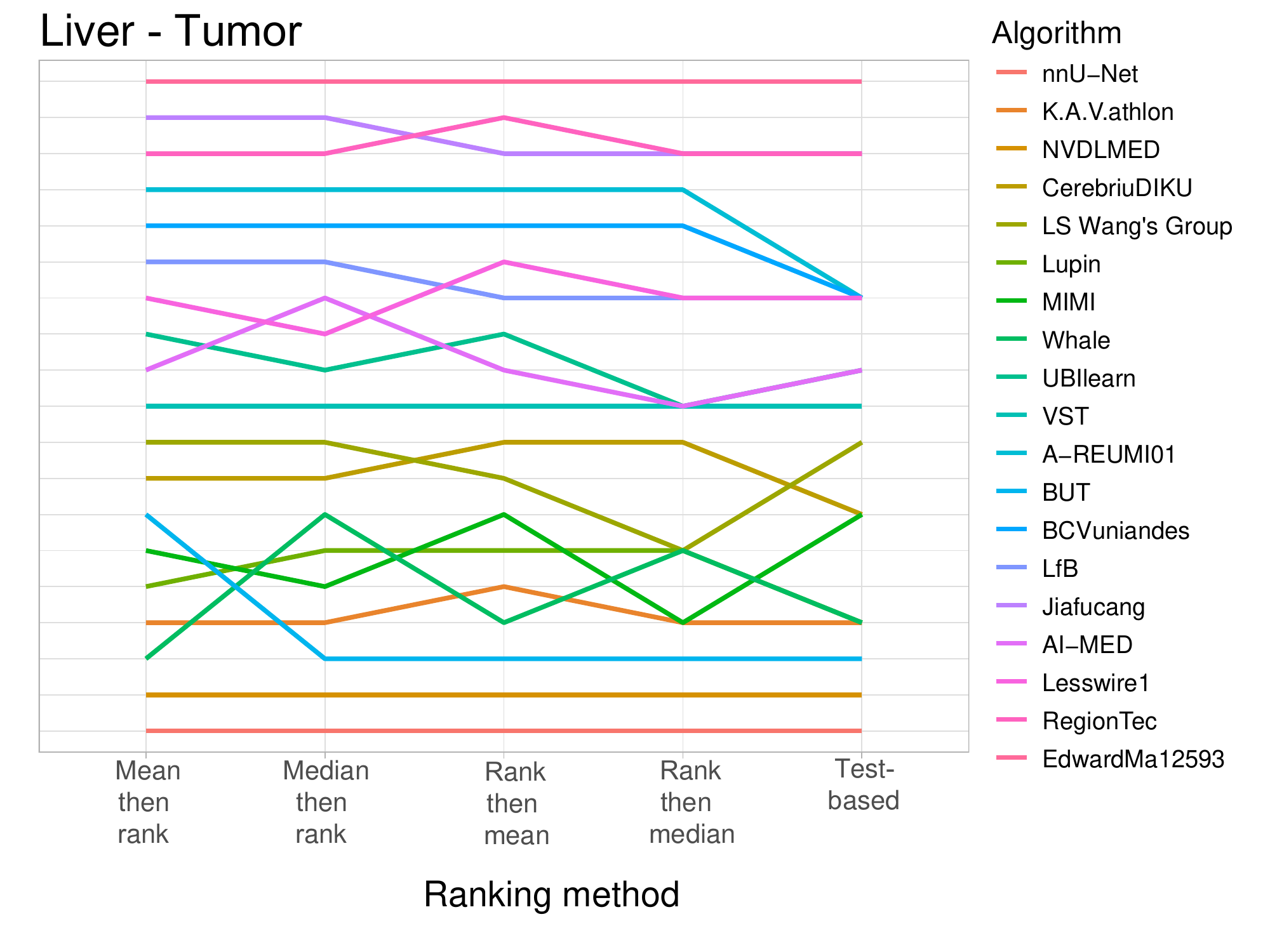}
    \end{minipage}
    \caption{Line plots visualizing rankings robustness across different ranking methods for the \textbf{liver} task. Each algorithm is represented by one colored line. For each ranking method encoded on the x-axis, the height of the line represents the corresponding rank. Horizontal lines indicate identical ranks for all methods.}
    \label{fig:line-plots-liver}
\end{figure}

\begin{table}[H]
\scriptsize
\centering
\caption{Mean Dice Similiarty Coefficient (DSC) values for all participating teams for all tasks (cancer) of the \textbf{lung} data set (the development phase).}
\label{tab:meanDSC-subtasks-lung-ph1}
\begin{center}
\begin{tabular}{lc}
\toprule
\textbf{Algorithm} & \textbf{Cancer} \\
\midrule
\textit{nnU-Net} & 0.69 \\
\textit{NVDLMED} & 0.52 \\
\textit{K.A.V.athlon} & 0.61 \\  
\textit{LS Wang's Group} & 0.55 \\
\textit{MIMI} & 0.55 \\
\textit{CerebriuDIKU}& 0.58 \\
\textit{Whale} & 0.51\\
\textit{UBIlearn}& 0.51\\
\textit{Lupin} & 0.55 \\
 \textit{Jiafucang} & 0.56 \\
\textit{LfB} & 0.47 \\
\textit{A-REUMI01}& 0.45 \\
\textit{VST} & 0.48\\
\textit{AI-Med}& 0.19\\
\textit{Lesswire1} & 0.18\\
\textit{BUT}& 0.33 \\
\textit{RegionTec} & 0.00 \\
\textit{BCVuniandes} & 0.56 \\
\textit{EdwardMa12593}& 0.08\\
\midrule
\textbf{Median} & \textbf{0.51} \\
\bottomrule
\end{tabular}
\end{center}
\end{table} 

\begin{figure}[H]
    \centering
    \includegraphics[width=0.6\textwidth]{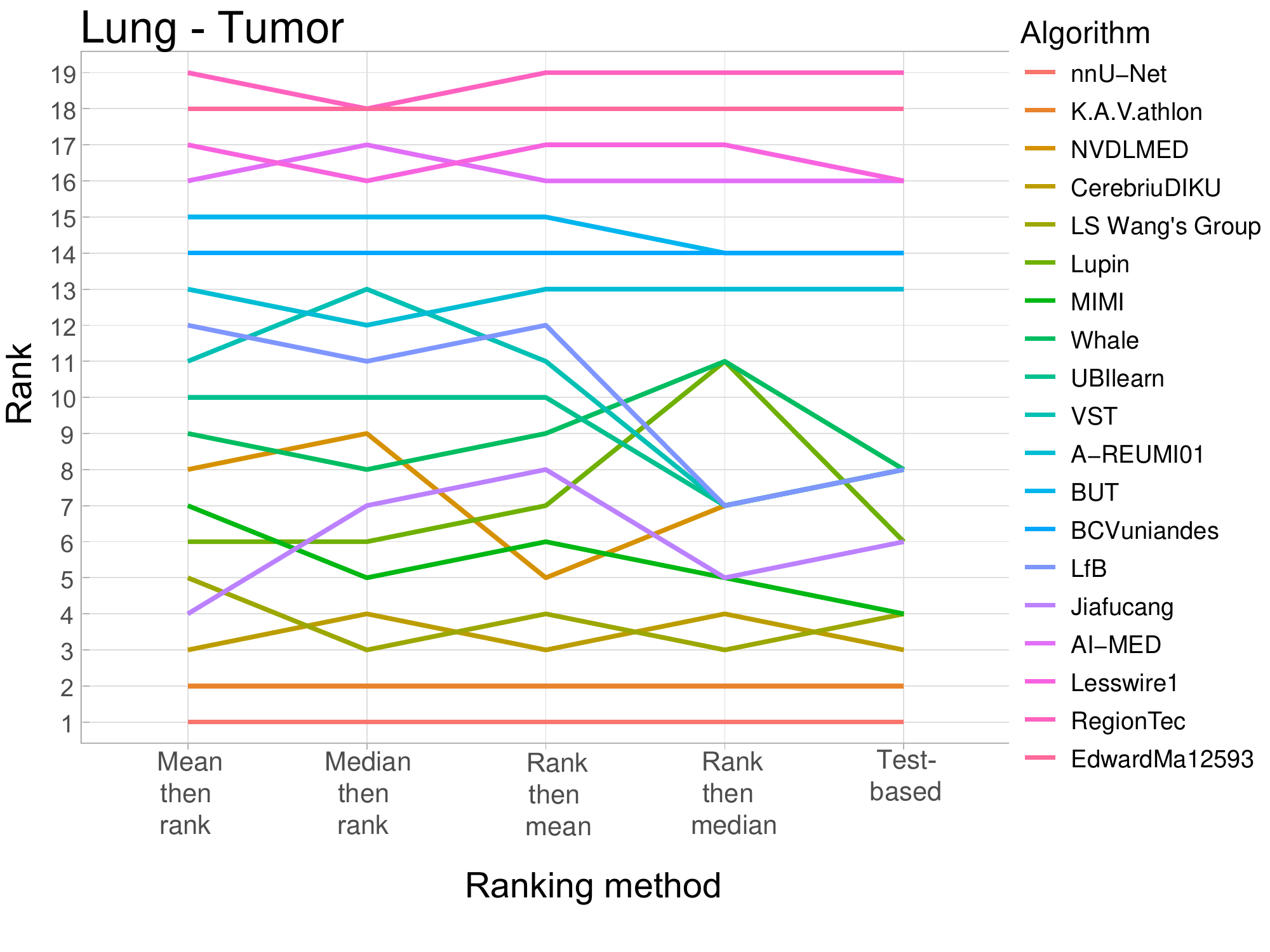}
    \caption{Line plots visualizing rankings robustness across different ranking methods for the \textbf{lung} task. Each algorithm is represented by one colored line. For each ranking method encoded on the x-axis, the height of the line represents the corresponding rank. Horizontal lines indicate identical ranks for all methods.}
    \label{fig:line-plots-lung}
\end{figure}

\begin{table}[H]
\scriptsize
\centering
\caption{Mean Dice Similiarty Coefficient (DSC) values for all participating teams for all tasks (pancreas, mass) of the \textbf{pancreas} data set (the development phase).}
\label{tab:meanDSC-subtasks-pancreas-ph1}
\begin{center}
\begin{tabular}{lcc}
\toprule
\textbf{Algorithm} & \textbf{Pancreas} & \textbf{Mass} \\
\midrule
\textit{nnU-Net} & 0.79 & 0.52 \\
\textit{NVDLMED} & 0.78 & 0.38 \\
\textit{K.A.V.athlon}& 0.75 & 0.43 \\  
\textit{LS Wang's Group} & 0.71 & 0.26\\
\textit{MIMI}& 0.70 & 0.26 \\
\textit{CerebriuDIKU} & 0.71 & 0.25\\
\textit{Whale} & 0.66 & 0.28\\
\textit{UBIlearn}& 0.69 & 0.18\\
\textit{Lupin}& 0.76 & 0.21\\
 \textit{Jiafucang} & 0.56 & 0.19 \\
\textit{LfB} & 0.59 & 0.28 \\
\textit{A-REUMI01}& 0.65 & 0.19 \\
\textit{VST} & 0.71 & 0.37 \\
\textit{AI-Med} & 0.48 & 0.04\\
\textit{Lesswire1} & 0.55 & 0.07 \\
\textit{BUT} & 0.73 & 0.21 \\
\textit{RegionTec}& 0.61 & 0.07 \\
\textit{BCVuniandes} & 0.56 & 0.15 \\
\textit{EdwardMa12593} & 0.50 & 0.01\\
\midrule
\textbf{Median} & \textbf{0.69} & \textbf{0.21} \\
\bottomrule
\end{tabular}
\end{center}
\end{table} 

\begin{figure}[H]
\hspace{-2cm}
    \centering
    \begin{minipage}[t]{0.55\textwidth}
    \includegraphics[width=1.2\textwidth]{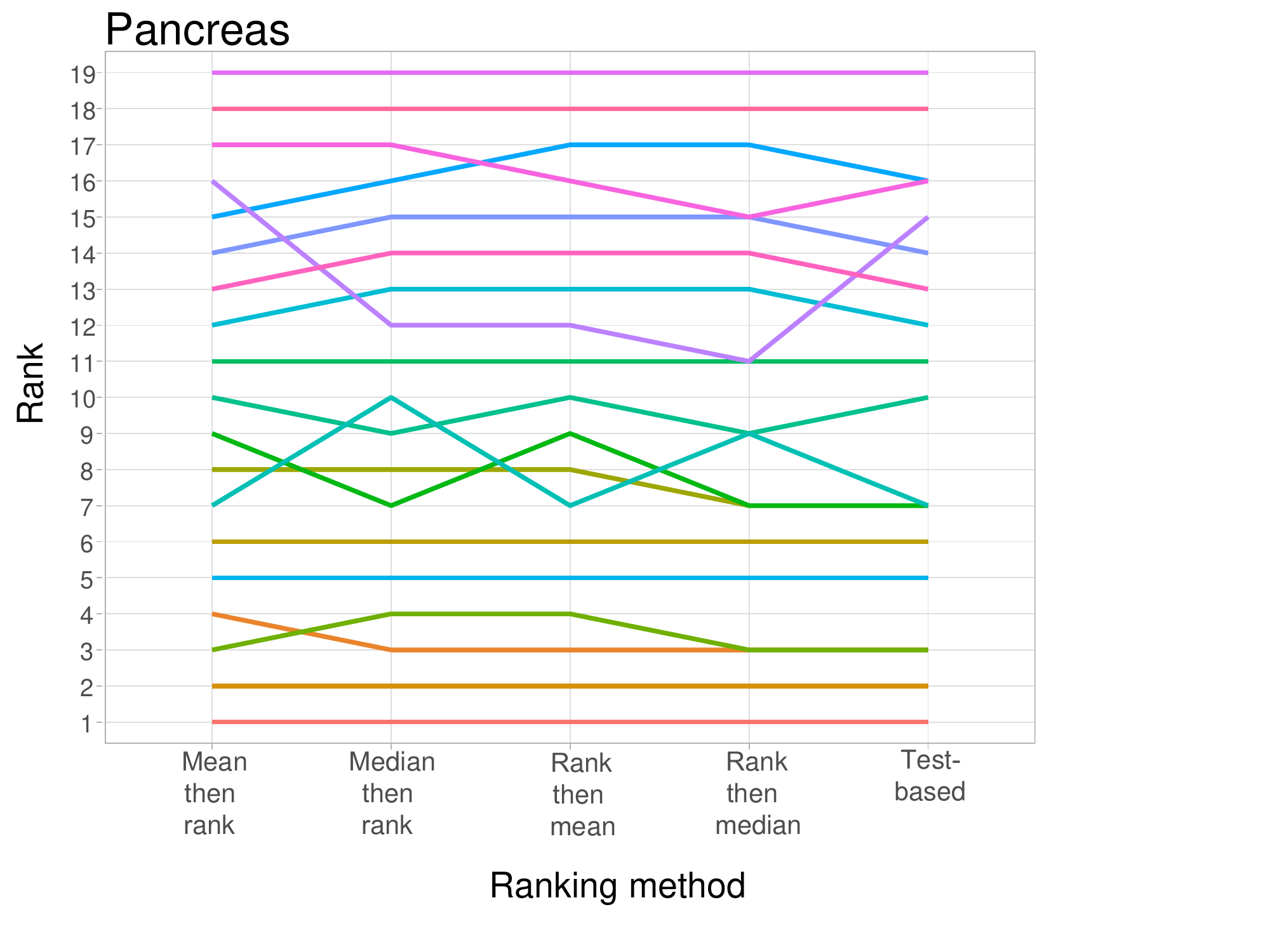}
    \end{minipage}
    \begin{minipage}[t]{0.55\textwidth}
    \includegraphics[width=1.2\textwidth]{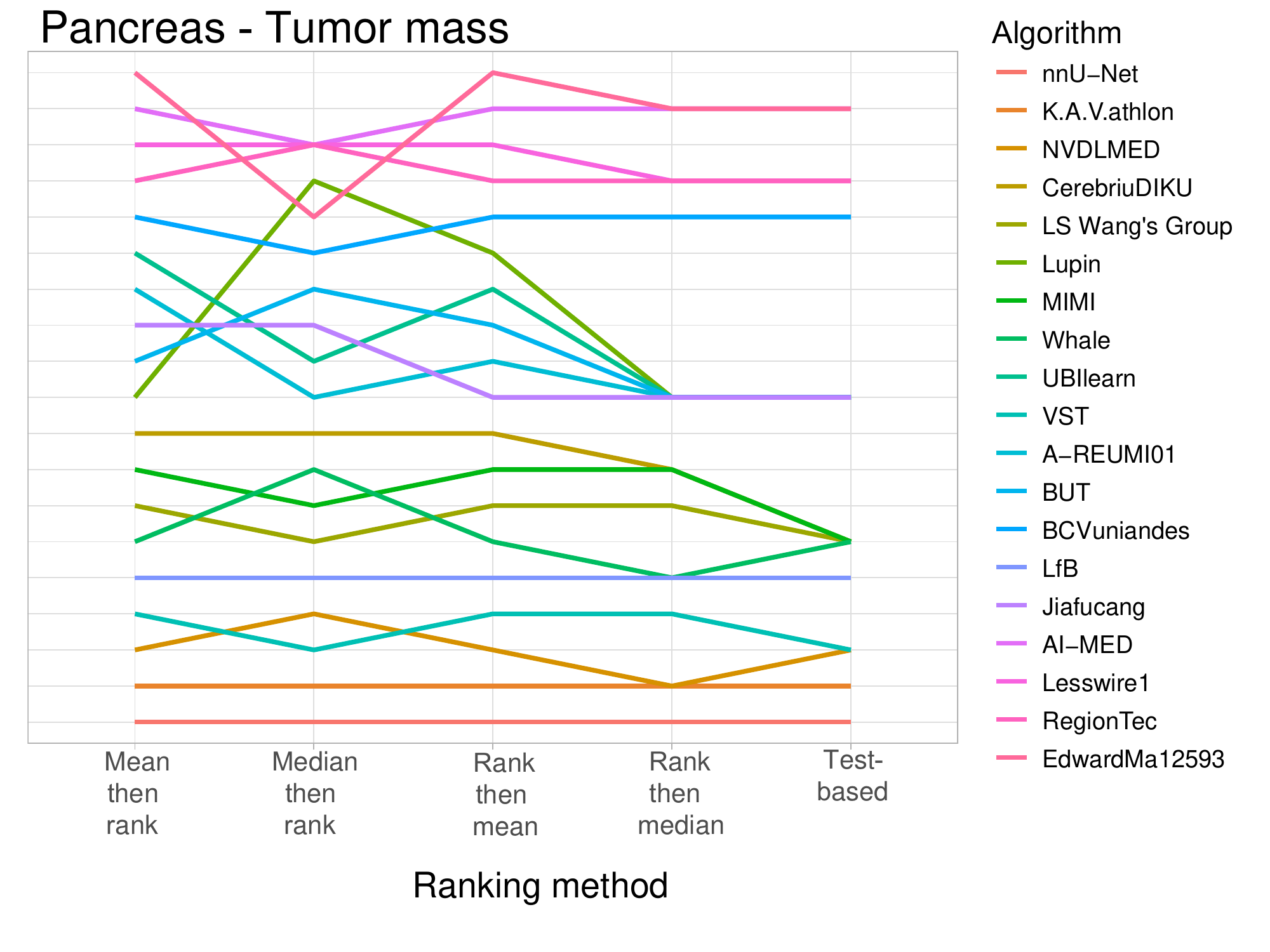}
    \end{minipage}
    \caption{Line plots visualizing rankings robustness across different ranking methods for the \textbf{pancreas} task. Each algorithm is represented by one colored line. For each ranking method encoded on the x-axis, the height of the line represents the corresponding rank. Horizontal lines indicate identical ranks for all methods.}
    \label{fig:line-plots-pancreas}
\end{figure}

\begin{table}[H]
\scriptsize
\centering
\caption{Mean Dice Similiarty Coefficient (DSC) values for all participating teams for all tasks (PZ, TZ) of the \textbf{prostate} data set (the development phase).}
\label{tab:meanDSC-subtasks-prostate-ph1}
\begin{center}
\begin{tabular}{lcc}
\toprule
\textbf{Algorithm} & \textbf{PZ} & \textbf{TZ} \\
\midrule
\textit{nnU-Net} & 0.76 & 0.89 \\
\textit{NVDLMED}& 0.69 & 0.87 \\
\textit{K.A.V.athlon}& 0.73 & 0.88\\  
\textit{LS Wang's Group} & 0.71 & 0.85\\
\textit{MIMI}&  0.71 & 0.87\\
\textit{CerebriuDIKU}& 0.69 & 0.86\\
\textit{Whale} & 0.70 & 0.88\\
\textit{UBIlearn} & 0.67 & 0.84\\
\textit{Lupin}& 0.72 & 0.88\\
 \textit{Jiafucang} & 0.70 & 0.84\\
\textit{LfB} & 0.58 & 0.82\\
\textit{A-REUMI01}& 0.67 & 0.86\\
\textit{VST} & 0.72 & 0.86 \\
\textit{AI-Med} & 0.56 & 0.78\\
\textit{Lesswire1} & 0.52 & 0.78\\
\textit{BUT}& 0.65 & 0.85\\
\textit{RegionTec} & 0.34 & 0.64\\
\textit{BCVuniandes} & 0.69 & 0.86\\
\textit{EdwardMa12593}& 0.43 & 0.74\\
\midrule
\textbf{Median} & \textbf{0.69} & \textbf{0.86} \\
\bottomrule
\end{tabular}
\end{center}
\end{table} 

\begin{figure}[H]
\hspace{-2cm}
    \centering
    \begin{minipage}[t]{0.55\textwidth}
    \includegraphics[width=1.2\textwidth]{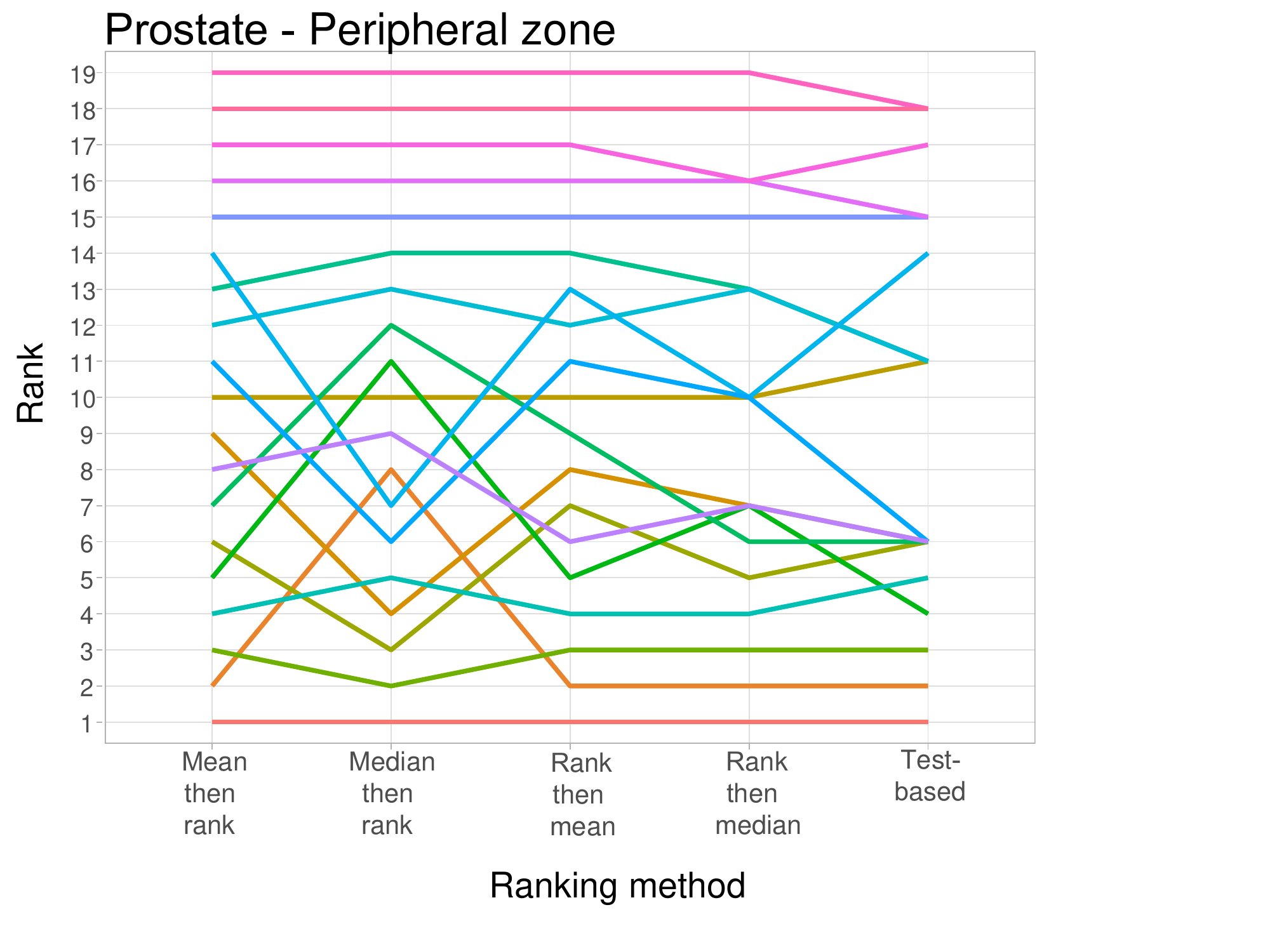}
    \end{minipage}
    \begin{minipage}[t]{0.55\textwidth}
    \includegraphics[width=1.2\textwidth]{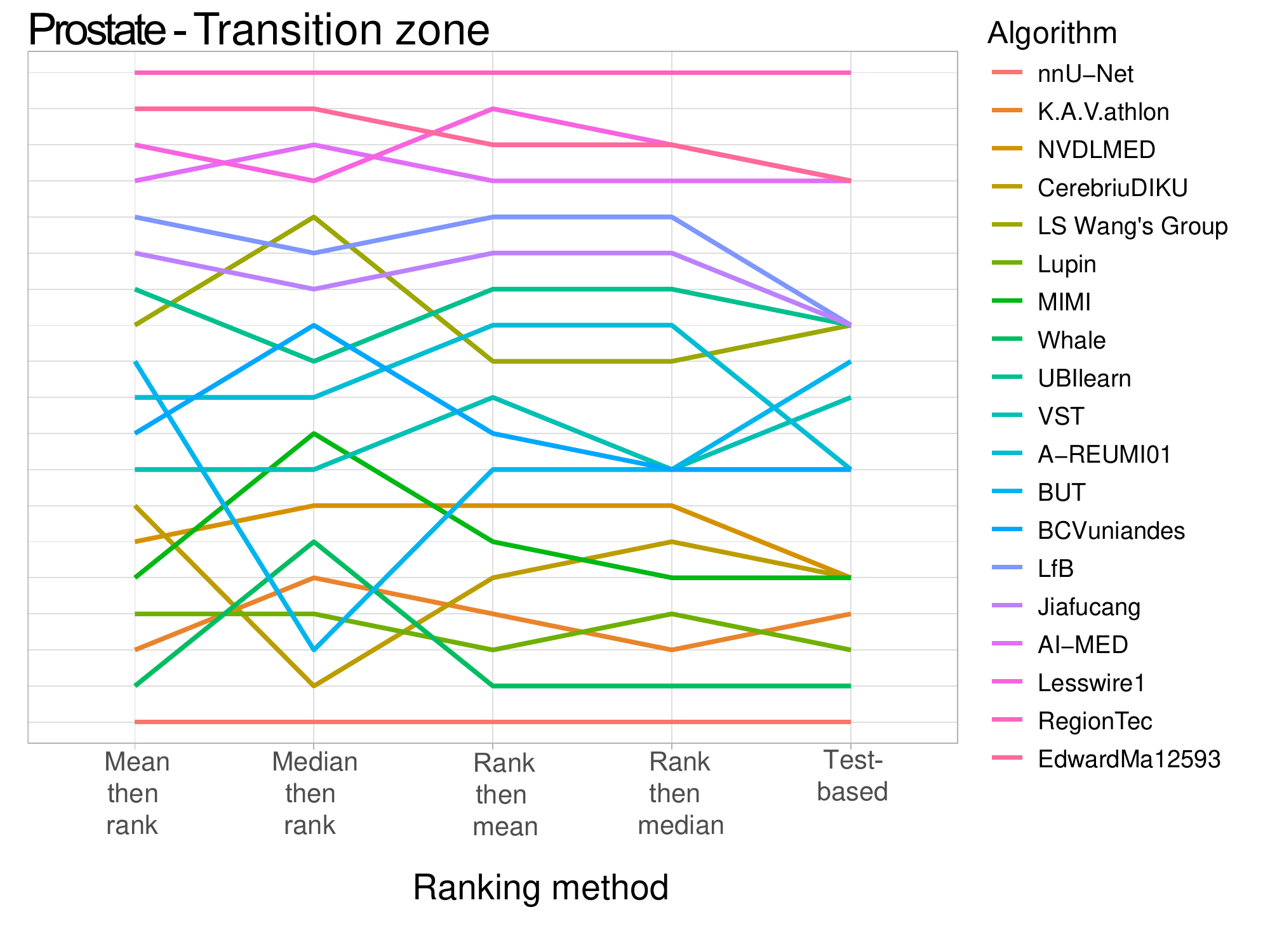}
    \end{minipage}
    \caption{Line plots visualizing rankings robustness across different ranking methods for the \textbf{prostate} task. Each algorithm is represented by one colored line. For each ranking method encoded on the x-axis, the height of the line represents the corresponding rank. Horizontal lines indicate identical ranks for all methods.}
    \label{fig:line-plots-prostate}
\end{figure}

\begin{table}[H]
\scriptsize
\centering
\caption{Mean Dice Similiarty Coefficient (DSC) values for all participating teams for all tasks (cancer primaries) of the \textbf{colon} data set (the mystery phase).}
\label{tab:meanDSC-subtasks-colon-ph2}
\begin{center}
\begin{tabular}{lc}
\toprule
\textbf{Algorithm} & \textbf{Cancer primaries} \\
\midrule
\textit{nnU-Net} & 0.56\\
\textit{NVDLMED}& 0.55\\
\textit{K.A.V.athlon}& 0.36 \\  
\textit{LS Wang's Group} & 0.41\\
\textit{MIMI}& 0.29 \\
\textit{CerebriuDIKU}& 0.28 \\
\textit{Whale} & 0.18\\
\textit{UBIlearn}&0.16 \\
\textit{Lupin}& 0.09 \\
 \textit{Jiafucang} & 0.19\\
\textit{LfB}& 0.24 \\
\textit{A-REUMI01}& 0.12 \\
\textit{VST}& 0.15 \\
\textit{AI-Med}& 0.11 \\
\textit{Lesswire1} & 0.09 \\
\textit{BUT}& 0.05 \\
\textit{RegionTec}& 0.06 \\
\textit{BCVuniandes} & 0.06 \\
\textit{EdwardMa12593}& 0.06 \\
\midrule
\textbf{Median} & \textbf{0.16} \\
\bottomrule
\end{tabular}
\end{center}
\end{table} 

\begin{figure}[H]
    \centering
    \includegraphics[width=0.6\textwidth]{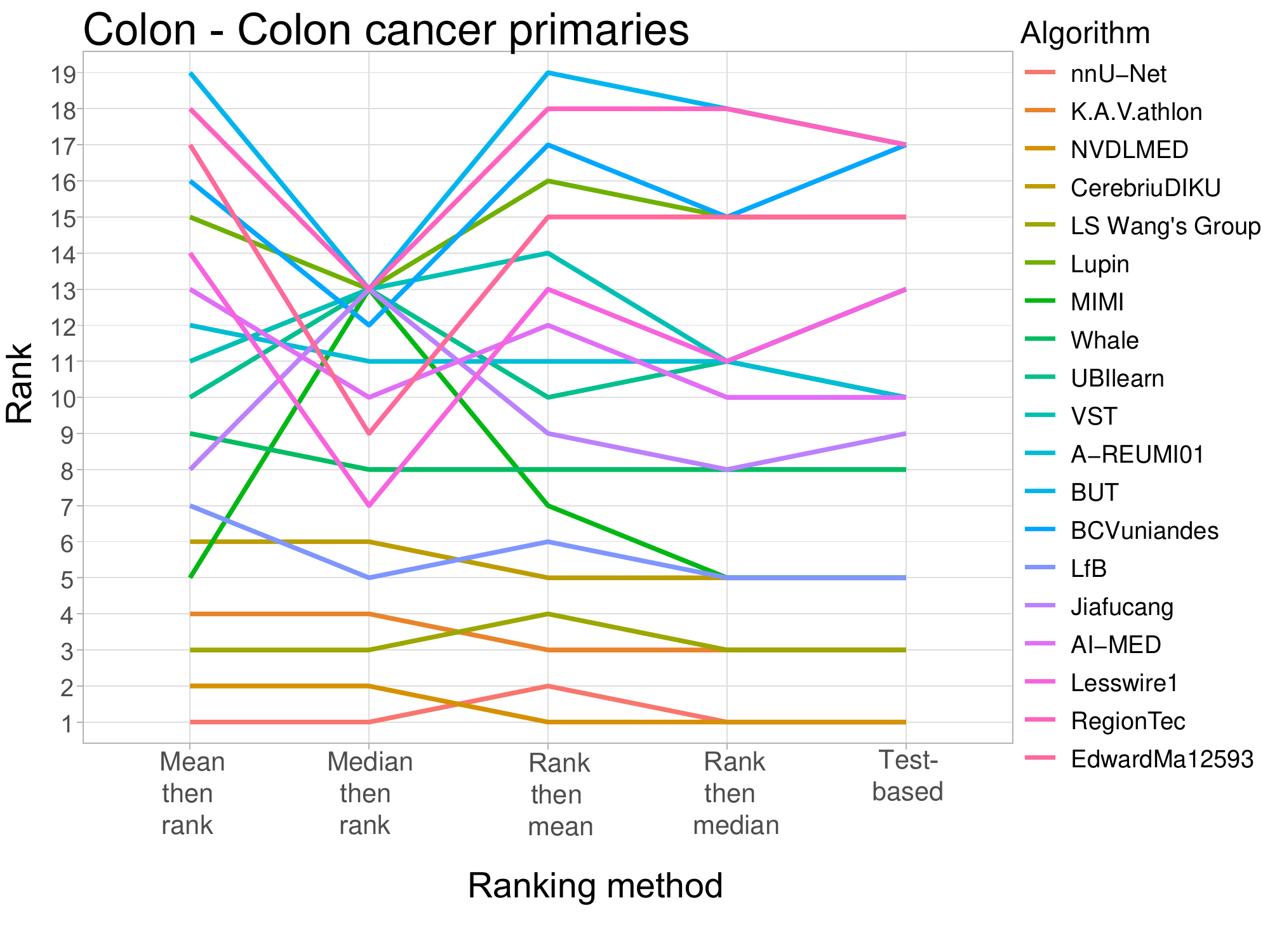}
    \caption{Line plots visualizing rankings robustness across different ranking methods for the \textbf{colon} task. Each algorithm is represented by one colored line. For each ranking method encoded on the x-axis, the height of the line represents the corresponding rank. Horizontal lines indicate identical ranks for all methods.}
    \label{fig:line-plots-colon}
\end{figure}

\begin{table}[H]
\scriptsize
\centering
\caption{Mean Dice Similiarty Coefficient (DSC) values for all participating teams for all tasks (vessel, tumor) of the \textbf{hepatic vessel} data set (the mystery phase).}
\label{tab:meanDSC-subtasks-hepvessel-ph2}
\begin{center}
\begin{tabular}{lcc}
\toprule
\textbf{Algorithm} & \textbf{Vessel} & \textbf{Tumor}  \\
\midrule
\textit{nnU-Net} & 0.63 & 0.69 \\
\textit{NVDLMED} & 0.63 & 0.64 \\
\textit{K.A.V.athlon} & 0.62 & 0.63 \\  
\textit{LS Wang's Group} & 0.55 & 0.64 \\
\textit{MIMI} & 0.59 & 0.56 \\
\textit{CerebriuDIKU} & 0.59 & 0.38 \\
\textit{Whale} & 0.56 & 0.46 \\
\textit{UBIlearn} & 0.59 & 0.37 \\
\textit{Lupin} & 0.59 & 0.47 \\
 \textit{Jiafucang} & 0.51 & 0.37\\
\textit{LfB} & 0.55 & 0.35 \\
\textit{A-REUMI01} & 0.56 & 0.39 \\
\textit{VST} & 0.44 & 0.36 \\
\textit{AI-Med} & 0.42 & 0.26 \\
\textit{Lesswire1} & 0.44 & 0.31 \\
\textit{BUT} & 0.44 & 0.39 \\
\textit{RegionTec} & 0.42 & 0.19 \\
\textit{BCVuniandes} & 0.14 & 0.32 \\
\textit{EdwardMa12593} & 0.14 & 0.11 \\
\midrule
\textbf{Median} & \textbf{0.55} &\textbf{0.38}\\
\bottomrule
\end{tabular}
\end{center}
\end{table} 

\begin{figure}[H]
\hspace{-2cm}
    \centering
    \begin{minipage}[t]{0.55\textwidth}
    \includegraphics[width=1.2\textwidth]{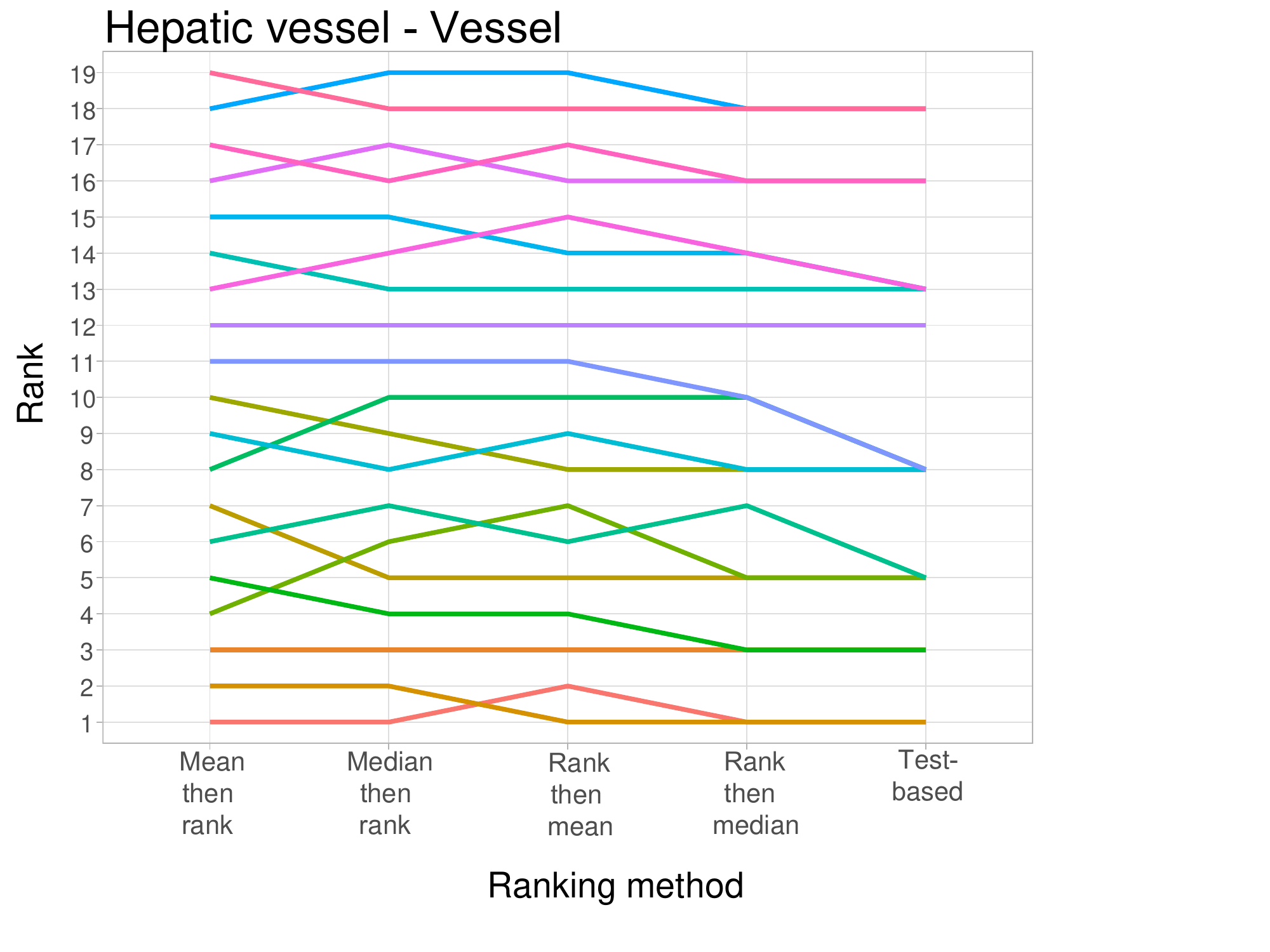}
    \end{minipage}
    \begin{minipage}[t]{0.55\textwidth}
    \includegraphics[width=1.2\textwidth]{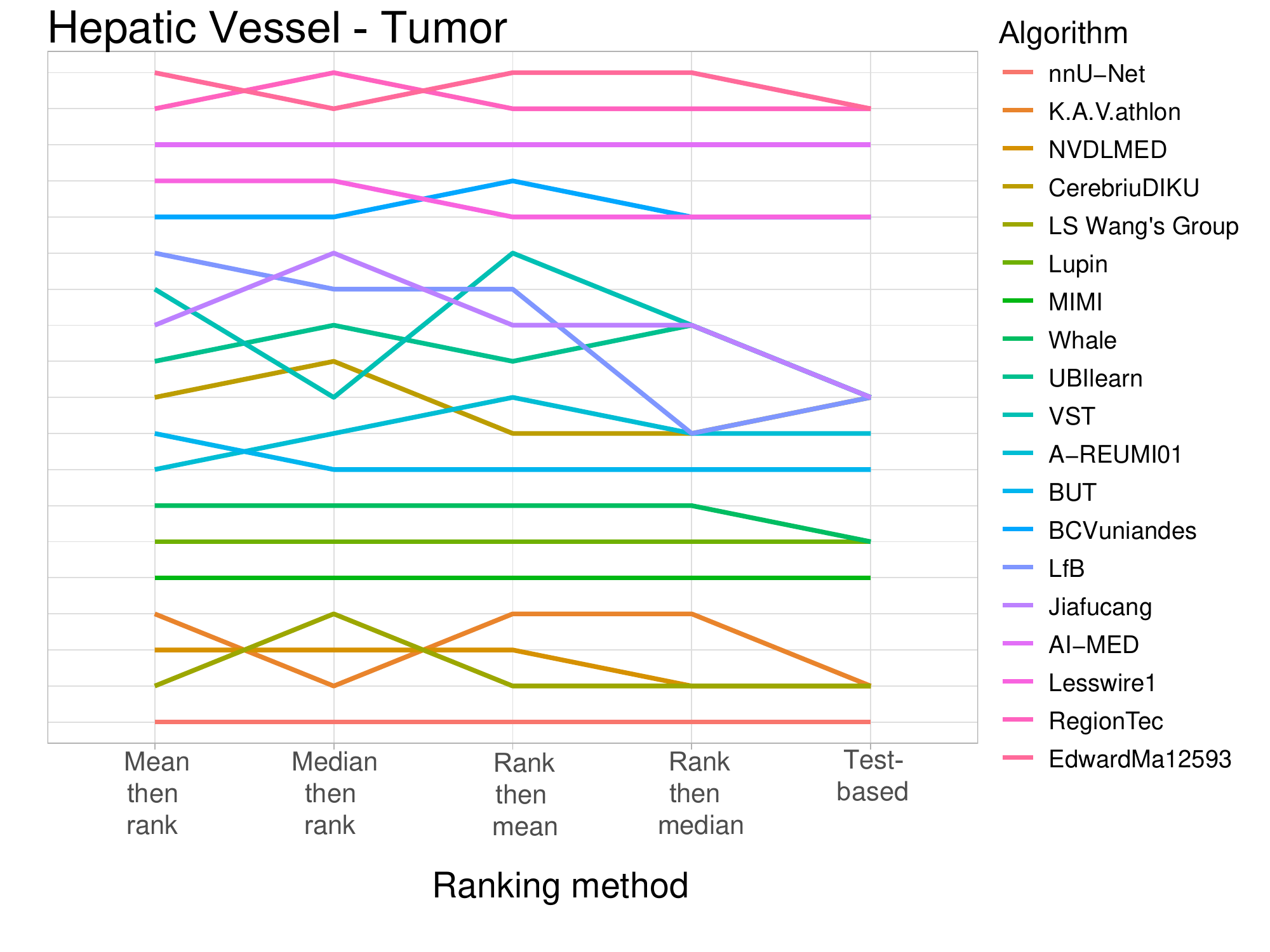}
    \end{minipage}
    \caption{Line plots visualizing rankings robustness across different ranking methods for the \textbf{hepatic vessel} task. Each algorithm is represented by one colored line. For each ranking method encoded on the x-axis, the height of the line represents the corresponding rank. Horizontal lines indicate identical ranks for all methods.}
    \label{fig:line-plots-hv}
\end{figure}

\begin{table}[H]
\scriptsize
\centering
\caption{Mean Dice Similiarty Coefficient (DSC) values for all participating teams for all tasks (spleen) of the \textbf{spleen} data set (the mystery phase).}
\label{tab:meanDSC-subtasks-spleen-ph2}
\begin{center}
\begin{tabular}{lc}
\toprule
\textbf{Algorithm} & \textbf{Spleen} \\
\midrule
\textit{nnU-Net} & 0.96\\
\textit{NVDLMED}& 0.96 \\
\textit{K.A.V.athlon}& 0.97\\  
\textit{LS Wang's Group} & 0.96\\
\textit{MIMI} & 0.93\\
\textit{CerebriuDIKU}& 0.95\\
\textit{Whale} & 0.95\\
\textit{UBIlearn}& 0.95\\
\textit{Lupin}& 0.94\\
 \textit{Jiafucang} & 0.93\\
\textit{LfB}& 0.83\\
\textit{A-REUMI01}&0.92\\
\textit{VST}& 0.94 \\
\textit{AI-Med}&  0.91\\
\textit{Lesswire1} & 0.86\\
\textit{BUT} & 0.89\\
\textit{RegionTec} & 0.92\\
\textit{BCVuniandes}& 0.82\\
\textit{EdwardMa12593}& 0.83\\
\midrule
\textbf{Median} & \textbf{0.94}\\
\bottomrule
\end{tabular}
\end{center}
\end{table} 

\begin{figure}[H]
    \centering
    \includegraphics[width=0.6\textwidth]{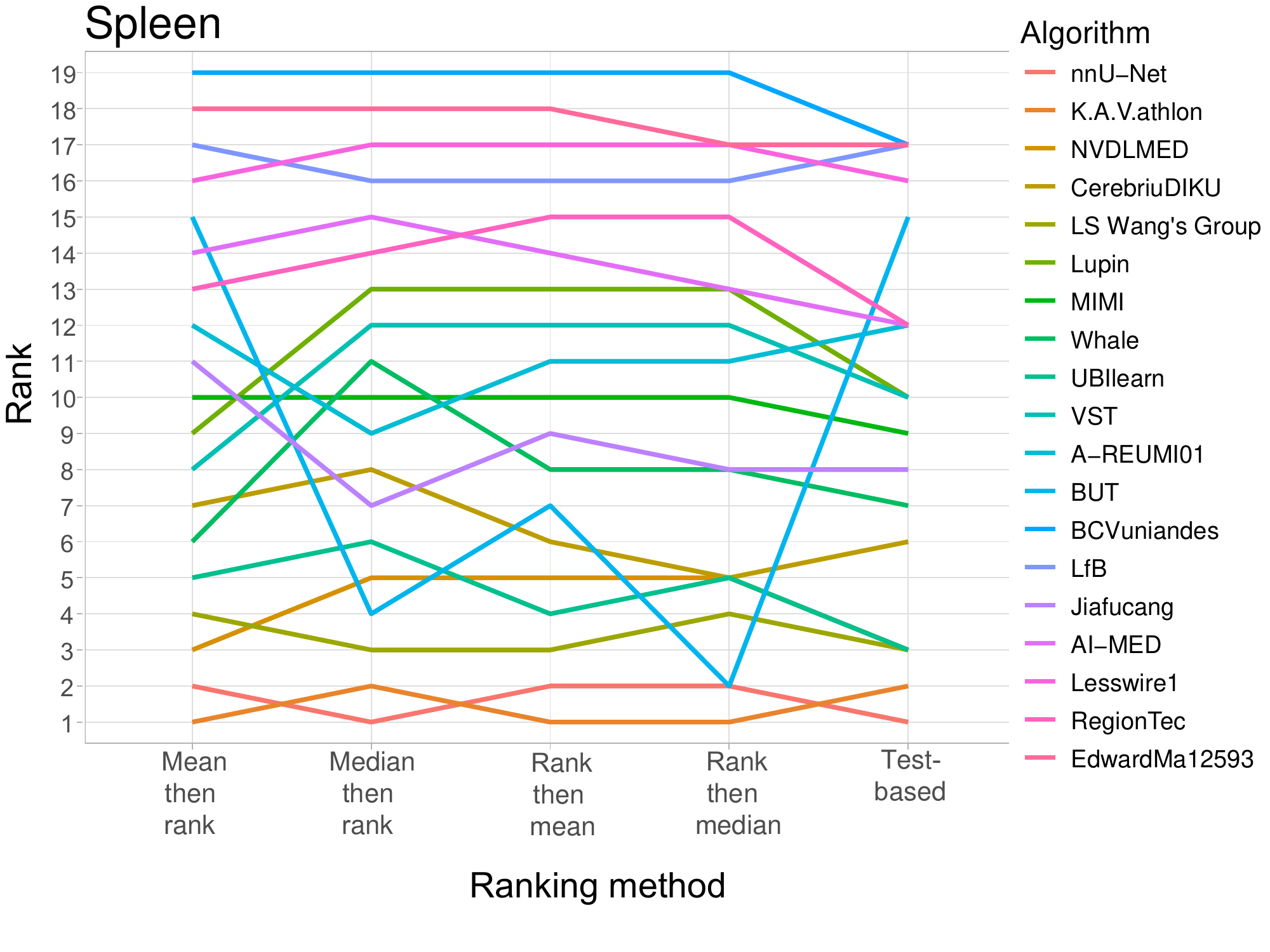}
    \caption{Line plots visualizing rankings robustness across different ranking methods for the \textbf{spleen} task. Each algorithm is represented by one colored line. For each ranking method encoded on the x-axis, the height of the line represents the corresponding rank. Horizontal lines indicate identical ranks for all methods.}
    \label{fig:line-plots-spleen}
\end{figure}

\newpage
\section{Bootstrap ranking analysis}
\begin{figure}[H]
    \makebox[\linewidth]{
        \includegraphics[width=1\textwidth]{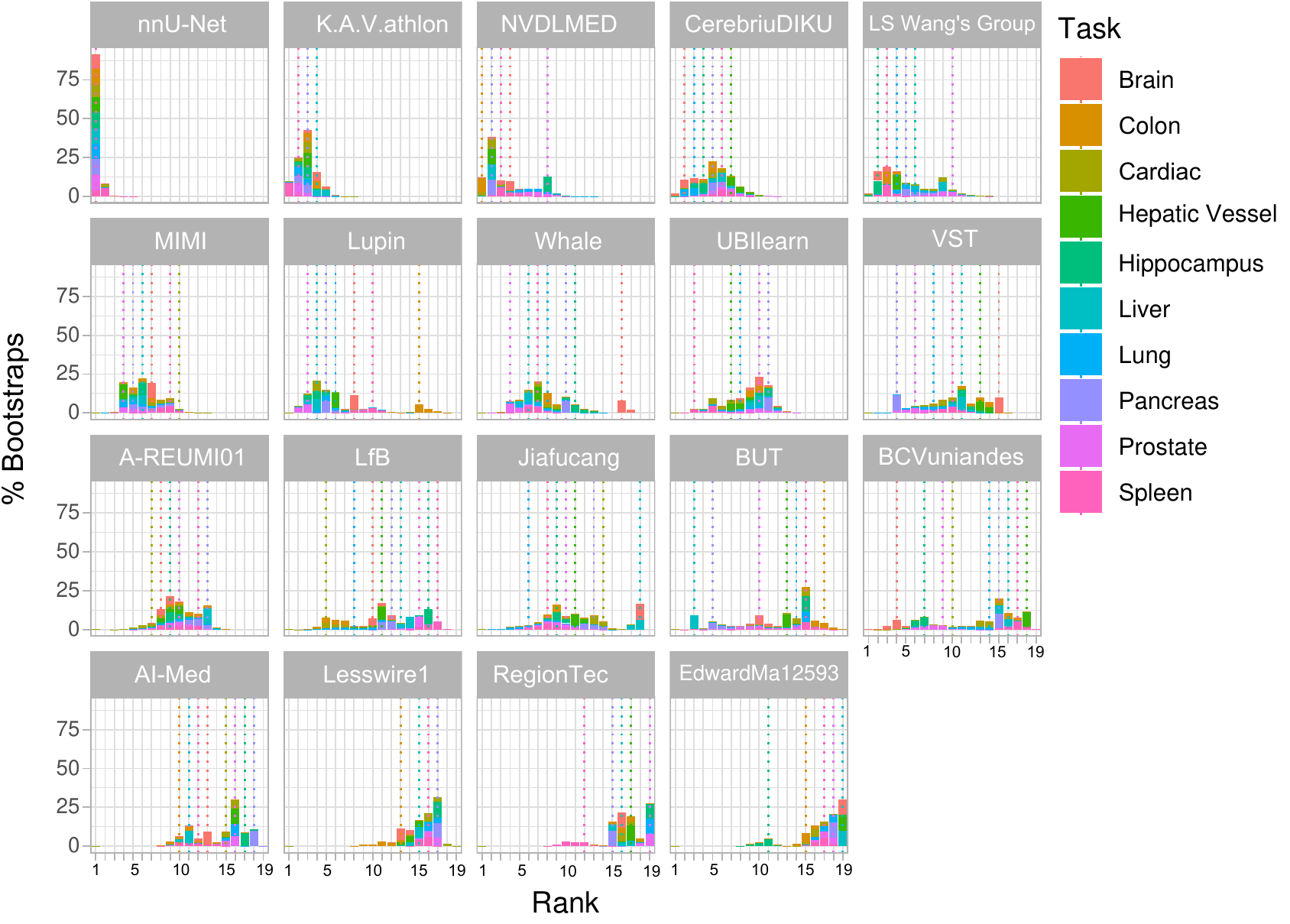}
    }
    \caption{Stacked frequency plot showing the achieved ranks of participating algorithms over 1,000 bootstrap datasets for all tasks (color-coded) for the DSC. Vertical lines indicate algorithms that achieved the same rank for the whole data set. The plot was created using \textit{challengeR}.}
    \label{fig:stackedplots}
\end{figure}

\section{Mean DSC values for the 2018 MSD challenge and the live-decathlon challenge}
\begin{figure}[H]
\hspace*{-2cm}
    \centering
    \includegraphics[width=1.2\textwidth]{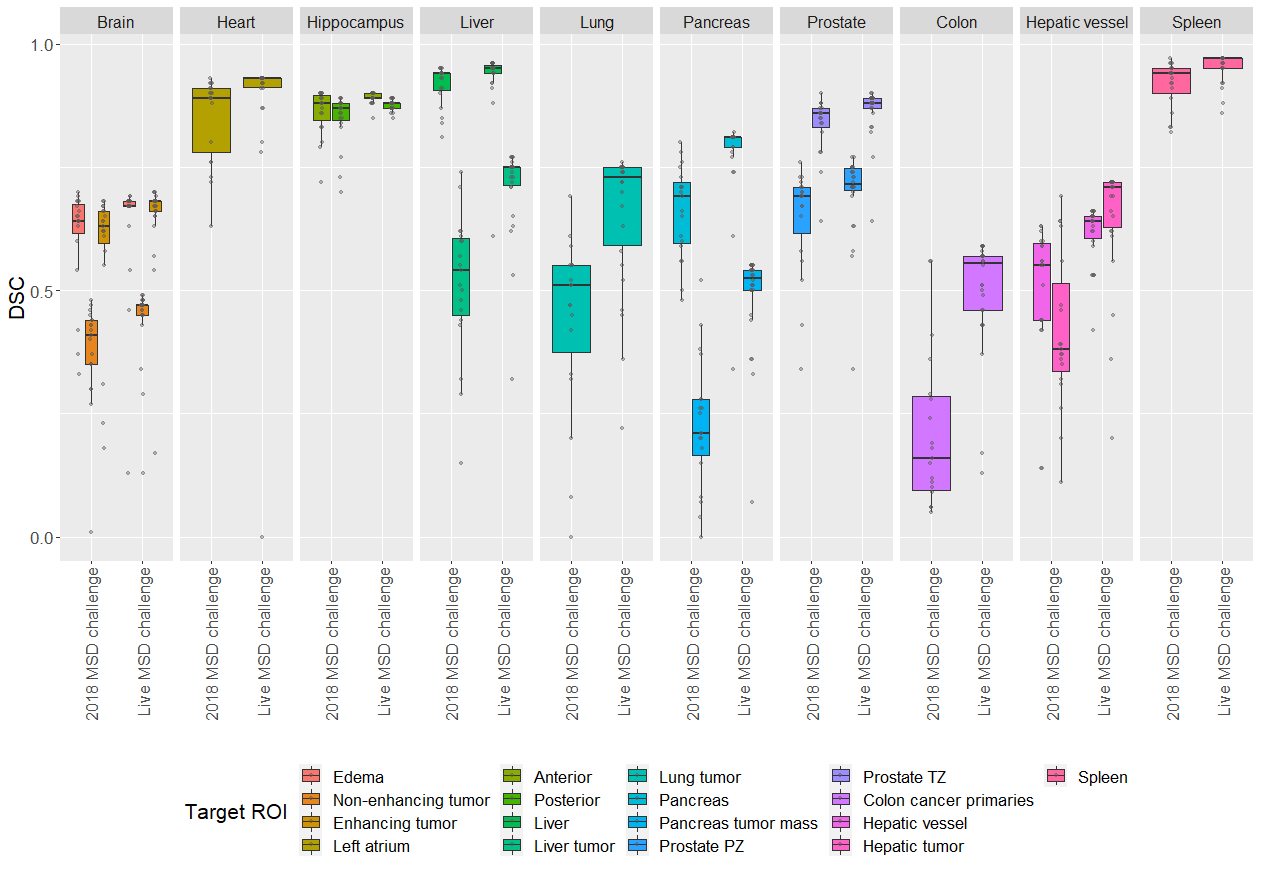}
    \caption{Dot- and box-plot of the mean DSC values computed for each task and target ROI for all algorithms in the 2018 MSD and live-decathlon challenges. box-plots represent descriptive statistics over all mean DSC values of each participant. The median value is shown by the black horizontal line within the box, the first and third quartiles as the lower and upper border of the box, respectively, and the 1.5 interquartile range by the vertical black lines. The mean DSC values per participant are provided as gray circles.}
    \label{fig:box-plot2018}
\end{figure}

\end{document}